\documentclass[11pt,draftcls,onecolumn]{IEEEtran}
\usepackage[dvips]{epsfig}
\usepackage[dvips]{graphicx}
\usepackage{boxedminipage}
\usepackage{color}
\usepackage{amsfonts}
\usepackage{amssymb}
\usepackage{amsmath}
\usepackage{graphics,amsmath,amsfonts,amssymb,epsfig,latexsym,amsfonts,graphicx,amssymb}
\usepackage{eqparbox}

\newcommand{\bfPhi}{{\mbox{\boldmath $\Phi$}}}

\newcommand{\bfSigma}{{\mbox{\boldmath $\Sigma$}}}

\newcommand{\bfPsi}{{\mbox{\boldmath $\Psi$}}}

\def\BibTeX{{\rm B\kern-.05em{\sc i\kern-.025em b}\kern-.08em
    T\kern-.1667em\lower.7ex\hbox{E}\kern-.125emX}}

\begin{document}
\title{Distributed Uplink Resource Allocation in
Cognitive Radio Networks -- Part I: Equilibria and Algorithms for
Power Allocation}
\author{Mingyi Hong, Alfredo Garcia and Stephen G. Wilson
\thanks{M. Hong and A. Garcia are with the Department of Systems and Information Engineering,
University of Virginia, Charlottesville, VA. Stephen Wilson is with
the Department of Electrical and Computer Engineering, University of
Virginia, Charlottesville, VA.} \thanks{Part of this manuscript has
been accepted by the Proceedings of IEEE INFOCOM 2011
\cite{hong11_infocom}.}}

\maketitle
\begin{abstract}
Spectrum management has been identified as a crucial step towards
enabling the technology of a cognitive radio network (CRN). Most of
the current works dealing with spectrum management in the CRN focus
on a single task of the problem, e.g., spectrum sensing, spectrum
decision, spectrum sharing or spectrum mobility. In this two-part
paper, we argue that for certain network configurations, jointly
performing several tasks of the spectrum management improves the
spectrum efficiency. Specifically, our aim is to study the uplink
resource management problem in a CRN where there exist multiple
cognitive users (CUs) and access points (APs). The CUs, in order to
maximize their uplink transmission rates, have to associate to a
suitable AP (spectrum decision), and to share the channels used by
this AP with other CUs (spectrum sharing). These tasks are clearly
interdependent, and the problem of how they should be carried out
efficiently and in a distributed manner is still open in the
literature.

In this first part of the paper, we focus on the problem of spectrum
sharing in a multi-channel CRN with a single AP. The insight gained
from the analysis of this simpler network is used as the building
block for analyzing the multiple AP network in the second part of
this paper. We formulate the single AP spectrum sharing problem into
a non-cooperative power allocation game, in which individual CUs aim
at maximizing their transmission rate by finding the suitable power
allocation on the available channels. Interestingly, we discover
that the set of equilibrium solutions of this game possesses the
following optimality properties: 1) any equilibrium solution is the
optimum input power allocation scheme in the sense that it maximizes
the sum rate of the network if joint decoding at the AP is employed;
2) asymptotically, when the number of channels becomes large, any
equilibrium solution becomes a Frequency Division Multiple Access
(FDMA) strategy, and the maximum system sum rate is achieved without
joint decoding. We subsequently propose a set of algorithms for the
CUs in the network to achieve such equilibrium solutions in
distributed fashion.
\end{abstract}
\section{Introduction}
\subsection{Motivation and Related Work}
The problem of distributed spectrum management in the context of
CRNs has been under intensive research recently. As pointed out by
the authors of \cite{akyildiz08}, spectrum management needs to
address four main tasks: 1) {\it spectrum sensing}, techniques that
ensure CUs to find the unused spectrum for communication; 2) {\it
spectrum decision}, protocols that enable the CUs to decide on the
best set of channels; 3) {\it spectrum sharing}, schemes that allow
different CUs to share the same set of channels; 4) {\it spectrum
mobility}, rules that require the CUs to leave the channel if
licensed users are detected. Many efforts have been devoted to
providing solutions to the individual tasks listed above. However,
as we will see in this two-part paper, in some CRN scenarios,
several of the above tasks become interdependent, and the CUs have
to perform these tasks {\it jointly} to achieve best performance. We
thus propose to provide solutions for the joint spectrum decision
and spectrum sharing problems in a multi-channel multi-user CRN.

In this two-part paper we focus on investigating an important CRN
configuration where such joint spectrum decision and spectrum
sharing is desirable. Consider a network with multiple CUs and APs.
The APs operate on non-overlapping spectrum bands, and the CUs need
to connect to one of the APs for communication. At this stage the
CUs essentially perform a spectrum decision task, in which they
decide on the best spectrum band to use, i.e., the best AP to
connect to. After the AP selection, the CUs can use { multiple
channels} belonging to the associated AP concurrently for
transmission, but different CUs interfere with each other if they
use the same channel. At this stage the CUs perform a spectrum
sharing task, in which multiple CUs use the same spectrum band for
communication. This network is a generalization of the single AP
network considered in previous literature, e.g., \cite{lai08},
\cite{Meshkati06} and \cite{islam08}. It also bears sufficient
similarity to the operational model of the IEEE 802.22 cognitive
radio standard \cite{stevenson09}, in which multiple service
providers install their respective APs to serve the same geographic
region. In the considered network, the CUs face the spectrum
decision problem when they select the AP, and they face the spectrum
sharing problem when they try to dynamically allocate their
communication power across the channels belonging to the selected
AP. Clearly, these two problems are strongly interdependent, as on
the one hand a particular CU has to select an AP before it can share
the spectrum assigned to this AP with all the other CUs associated
with it; on the other hand, after sharing the spectrum, an
individual CU may have the incentive to switch to a different AP if
it perceives that such action will increase its communication rate.
A poor spectrum decision and spectrum sharing scheme will not only
lead to unsatisfactory performance for individual CUs, but also
result in an unstable system in which CUs are constantly unsatisfied
with their current communication rates and consequently changing
their AP associations and power allocation indefinitely.

In the first part of this paper, we focus on the spectrum sharing
aspect of the above problem. Specifically, we study the uplink power
allocation problem in a CRN with a {\it single} AP. This problem is
important in its own right, and the insight gained from studying
this problem serves in studying the more complicated network with
multiple APs that we discuss in the second part of this paper.

Centralized strategies for resource allocation in multi-carrier
single AP network has been extensively studied. In \cite{song05a}
and \cite{song05b}, a joint sub-carrier assignment and power
allocation algorithm is proposed for downlink orthogonal frequency
division multiplexing (OFDM) network with the objective to optimize
utility functions related to throughput and fairness. It has been
shown that when the utility function is properly chosen, the optimum
access strategy is FDMA, and the downlink throughput can achieve
Shannon capacity. In \cite{luo04}, the authors formulate the optimum
(in the sense of minimizing the received mean square error) linear
transceiver design problem in a multiple access (MAC) intersymbol
interference (ISI) channel into an optimum uplink subcarrier
allocation and power loading problem, and propose a strongly
polynomial algorithm to determine such optimum strategy.
\cite{yu02b} proposes a numerical method to compute the capacity of
FDMA MAC channel as well as to (near-) optimally assign the
channels. The proposed method assigns the channel to different users
by solving a (convexly relaxed) optimization problem. \cite{kim05b}
and \cite{liu10} are two recent developments for uplink/downlink
resource allocation in multi-carrier systems. The uplink optimality
of OFDMA system has been discussed in \cite{Li07}, in which the
authors derived sufficient conditions of the channel state as well
as the received signal noise ratio for the OFDMA system to achieve
maximum uplink system sum rate. We note that the centralized schemes
usually assume that the AP carries out all the necessary
computations and enforces the resultant optimum policies among the
mobile users in the network.

However, such centralized scheme may not be applicable in networks
where individual mobile users are autonomous or selfish and have the
intention and the ability  to deviate from the centralized policies
(e.g., in the cognitive radio network). Consequently various
distributed algorithms are proposed in the literature, for example,
\cite{he08}, \cite{acharya09} and \cite{Meshkati06}. In \cite{he08},
a distributed power allocation scheme is proposed for uplink OFDM
systems where the channel state is simplified to having only
discrete levels. Notably, this scheme only requires that each user
has the knowledge of {\it its own} channel state information (CSI),
thus the signaling needed for the users to obtain the {\it global}
CSI (as required by the algorithms proposed in, say, \cite{lai08})
from the AP is greatly reduced. \cite{acharya09} is a recent work
considering the uplink dynamic spectrum sharing problem in a
multi-carrier multiple service provider cognitive network. The
authors developed a distributed algorithm for the users to jointly
choose the size of the spectrum as well as the amount of power for
transmission. However, in this work the channel is considered to be
{\it flat} for each user, as a result, the users only need to select
the {\it size} of spectrum they need (because any portions of the
spectrum of the same size is equivalent to the users), which greatly
simplifies the analysis.  In \cite{Meshkati06}, the problem of
distributed energy-efficient power control in uplink multi-carrier
CDMA system is considered. The authors formulate the problem into a
game-theoretic framework, and a distributed algorithm is proposed in
which each user transmits only on its ``best" channel. \cite{yu04}
considered a generalization of a multi-carrier MAC channel, and
proposed a distributed iterative water-filling (IWF) algorithm to
compute the maximum sum capacity of the system. It is worth
mentioning that the algorithms proposed in \cite{Meshkati06} and
\cite{yu04} both require that the users update their transmission
strategies {\it sequentially}, thus may result in slow convergence
when the number of users in the system is large.

In this first part of the paper, we formulate the uplink spectrum
sharing problem into a non-cooperative game, in which the CUs in the
network try to maximize their individual transmission rate. Due to
the structure of the considered multi-channel network, we are able
to identify the proposed game as a {\it potential game}
\cite{monderer96}, in which the players in the game, although
selfish by nature, behave as if they aim at jointly optimizing an
objective function (which is called the {\it potential function}).
Such underlying structure of the game allows us to characterize many
optimality properties of the equilibrium solution. In particular,
the maximum value of the potential function equals to the maximum
sum rate achievable for the network. As a result, the CUs can be
viewed as jointly optimizing the system sum rate. As far as we know,
such interesting relationship between a spectrum sharing game and
the system sum rate in multi-carrier single AP network has not been
shown in the literature. We then propose three algorithms with
convergence guarantees that allow the CUs to reach the equilibrium
solution(s) in a distributed fashion. For the sake of rigor, we
categorize our algorithms either as ``weak convergent", in which the
individual CUs' strategies converge to {\it the set} of equilibria,
or as ``strong convergent", in which the individual CUs' strategies
converge to an equilibrium point. Such distinction is technical yet
necessary, as we will point out in section \ref{subInapplicableIWF},
for the reason that unlike most other spectrum sharing games (e.g.,
\cite{scutari08a}, \cite{luo06b}), our game generally admits {\it a
connected set of equilibria}, consequently it is possible, at least
theoretically, for the algorithm to converge to the set of
equilibria without converging to any equilibrium point.

\subsection{Organization of This Work and Notations}
This part of the paper is organized as follows. In section
\ref{secProblemFormulation}, we present the system under
consideration, and formulate the spectrum sharing problem into a
non-cooperative game. In section \ref{secNEProperty}, we give
detailed analysis regarding to the property of the NE of the game.
In section \ref{secAlgorithm}, we propose distributed algorithms and
study their convergence properties. In section \ref{secSimulation},
we show the numerical result. This part of the paper concludes in
section \ref{secConclusion}.

Some notations used in this paper are specified as follows: we use
bold lowercase and uppercase letters for vectors and matrices,
respectively. The $(i,j)^{th}$ element of a matrix $\mathbf{X}$ is
denoted by $[\mathbf{X}]_{i,j}$. For a symmetric matrix
$\mathbf{X}$, $\mathbf{X}\succeq 0$ signifies that $\mathbf{X}$ is
positive semidefinite. The trace of a matrix is denoted by
$tr(\mathbf{X})$; the determinant of a matrix is denoted by
$|\mathbf{X}|$. $\mathbf{I}_n$ is used to denote a $n\times n$
identity matrix. For a vector $\mathbf{x}$, $diag(\mathbf{x})$
represents a diagonal matrix with its diagonal entries equal to the
entries of the vector $\mathbf{x}$. We use $\rho(\mathbf{X})$ to
denote the spectral radius of the matrix $\mathbf{X}$.


\section{Problem Formulation}\label{secProblemFormulation}
\subsection{System Model}
We consider a wireless network with a set
$\mathcal{N}\triangleq\{1,2,\cdots,N\}$ CUs and a single AP. Let us
normalize the total available bandwidth to 1, and divide it equally
into $K$ channels; let the set $\mathcal{K}\triangleq\{1,2,\cdots,K
\}$ represents the set of available channels.

The followings
are our main assumptions of the network.\\
{\bf A-1)} The available spectrum can be used exclusively by
the CRN, for a relative long period of time.\\
{\bf A-2)} Each CU can concurrently use all the channels of the AP
for transmission, if desired. \\
{\bf A-3)} The AP is equipped with single-user receivers.

Assumption A-1) can be achieved either under the spectrum property
right model in which the licensed networks sell or lease the
spectrum to the cognitive network for a period of time for exclusive
use, or under the situation that the cognitive network exploits
relatively static spectrum white spaces unused by local TV broadcast
\cite{zhao07}. Assumption A-3) is congruent with the lack of
coordination of the CUs, because an individual CU essentially treats
other CUs' transmission as noises. It also allows for implementation
of low-complexity receivers at the AP. This assumption is generally
accepted in designing distributed algorithm in multiple-access
channels \cite{lai08}.

Let $x_i(k)$ denote the complex Gaussian signal transmitted by CU
$i$ on channel $k$; let $p_i(k)=E[(x_i(k))^2]$ denote the
transmitted power of CU $i$ on channel $k$. Let $z(k)\sim CN(0,
n(k))$ denote the white complex Gaussian environment noise
experienced at the receiver of AP with mean zero and variance
$n(k)$. Let $h_i(k)$ denote the channel gain coefficient between CU
$i$ and the AP on channel $k$. The signal received at the AP on
channel $k$, denoted by $y(k)$, can then be expressed as:
\begin{align}
y(k)=\sum_{i=1}^{N}x_i(k) h_i(k)+z(k).
\end{align}
Define $\mathbf{x}_i=\left[x_i(1),\cdots,x_i(K)\right]^{\intercal}$;
$\mathbf{z}=\left[z(1),\cdots,z(K)\right]^{\intercal}$;
$\mathbf{H}_i=diag\left[h_i(1),\cdots,h_i(K)\right]$. Then the
received vector signal
$\mathbf{y}=\left[y(1),\cdots,y(K)\right]^{\intercal}$ can then be
expressed concisely as:
\begin{align}
\mathbf{y}=\sum_{i=1}^{N}\mathbf{H}_i\mathbf{x}_i+\mathbf{z}.\label{eqMultiChannelMAC}
\end{align}


Let $\mathbf{p}_i=\left[p_i(1),\cdots,p_i(K)\right]^{\intercal}$ be
CU $i$'s transmission power profile; let
$\mathbf{p}_{-i}=\left[\mathbf{p}^{\intercal}_1,\cdots,\mathbf{p}^{\intercal}_{i-1},\mathbf{p}^{\intercal}_{i+1},
\cdots,\mathbf{p}^{\intercal}_{N}\right]^{\intercal}$ be the
transmission profile of all other CUs except CU $i$; let
$\mathbf{p}=\left[\mathbf{p}^{\intercal}_1,\cdots,\mathbf{p}^{\intercal}_{N}\right]^{\intercal}$
be the system power profile. Define $\bar{p}_i$ to be CU $i$'s
maximum allowable transmission power, then its feasible space can be
expressed as
$\mathcal{P}_i\triangleq\left\{\mathbf{p}_i:\mathbf{p}_i\ge\mathbf{0},
\sum_{k=1}^{K}p_i(k)\le \bar{p}_i\right\}$. In this network when
single-user receiver is employed at the AP, and when assuming other
CUs' power profiles are fixed, the CU $i$'s maximum achievable rate
can be expressed as \cite{cover05}:
\begin{align}
R_i(\mathbf{p}_i,\mathbf{p}_{-i})=\frac{1}{K}\sum_{k=1}^{K}\log\left(1+\frac{{p}_i(k)|h_i(k)|^2}{n(k)+\sum_{j\ne
i}{p}_j(k)|h_j(k)|^2}\right).
\end{align}

We assume that each CU $i\in\mathcal{N}$ has the knowledge of its
own channel coefficients $\{h_i(k)\}_{k\in\mathcal{K}}$, and the
quantity $\sum_{j\ne i}p_j(k)|h_j(k)|^2+n(k)$ on every channel,
which represents the sum of noise plus interference on each channel.
This information can be obtained by the AP and fed back to each CU
$i$, as suggested in \cite{Meshkati06}. We do not assume that
individual CU has any information regarding the other CUs' channel
coefficients; nor do we require that individual CU $i$ knows the
power budget $\{\bar{p}_j\}_{j\ne i}$ or the power allocations
$\mathbf{p}_{-i}$ of other CUs.

The problem that the CUs are facing is that under the above sets of
system constraints, how should they decide on the policy for
efficiently sharing of the available spectrum in a distributed
manner? In the following subsection, we formulate such spectrum
sharing problem into a game-theoretical framework.

\subsection{A Non-Cooperative Game Formulation}
In order to facilitate the development of a distributed algorithm,
we model each CU as selfish agent, and its objective is to maximize
its own transmission rate. More specifically, when $\mathbf{p}_{-i}$
is fixed, CU $i$ is interested in solving the following optimization
problem:
\begin{align}
&\max_{\mathbf{p}_i\in\mathcal{P}_i}~\frac{1}{K}\sum_{k=1}^{K}\log\left(1+\frac{{p}_i(k)|h_i(k)|^2}
{n(k)+\sum_{j\ne i}{p}_j(k)|h_j(k)|^2}\right)\label{eqUserProblem}.
\end{align}

The solution to this optimization problem is the well-known
single-user water-filling solution \cite{cover05}:
\begin{align}
\Phi^k_i(\mathbf{p}_{-i})\triangleq\left[\sigma_i-\frac{n(k)+\sum_{j\ne
i}{p}_j(k)|h_j(k)|^2}{|h_i(k)|^2}\right]^{+},~\forall~k\in\mathcal{K}
\label{eqWaterFilling}
\end{align}
where $\sigma_i\ge 0$ is the dual variable for the sum-power
constraints.

We introduce a non-cooperative spectrum sharing game where the
players of the game are the CUs in the network, the utility of each
player is its achievable transmission rate, and the strategy of each
player is its transmit power profile. We denote this game as
$G=\{\mathcal{N},\mathcal{P},\{R_i(.)\}_{i\in\mathcal{N}}\}$, where
$\mathcal{P}=\prod_{i\in\mathcal{N}}\mathcal{P}_i$ is the joint
feasible region of all CUs.

The Nash Equilibrium (NE) of the above game is defined as the
strategies $\{\mathbf{p}_i^*\}_{i\in\mathcal{N}}$ satisfying
\cite{osborne94}:
\begin{align}
\mathbf{p}^{*}_i\in \arg_{\mathbf{p}_i\in\mathcal{P}_i}\max
R_i(\mathbf{p}_i,\mathbf{p}^*_{-i})~\forall~i\in\mathcal{N}.\label{eqDefineNE}
\end{align}

Intuitively, a NE of the game is a stable point of the system where
no player has the incentive to deviate from its current strategy. In
the following sections, we will first set out to analyze the
properties of the set of NE of game $G$, and then propose
distributed algorithms to reach the set of NE.

\section{Characteristics and Optimality of the NE}\label{secNEProperty}
\subsection{NE as Maximizers of the Potential
Function}\label{subNEPotential}

In order to facilitate the analysis, we introduce the notion of a
potential function of the game $G$. Define a concave function $P:
\mathcal{P}\to \mathbb{R}$:
\begin{align}
P(\mathbf{p})\triangleq
\frac{1}{K}\sum_{k=1}^{K}\left(\log\left(n(k)+\sum_{i=1}^{N}|h_i(k)|^2p_i(k)\right)-\log(n(k))\right).\label{eqPotential}
\end{align}

We can readily observe that the following identity is true for all
$i\in\mathcal{N}$ and $k\in\mathcal{K}$:
\begin{align}
\frac{\partial P(\mathbf{p})}{\partial{{p}_i}(k)}=\frac{\partial
R_i(\mathbf{p}_i,\mathbf{p}_{-i})}{{{p}_i}(k)},\label{eqPotentialProperty1}
\end{align}
or similarly, for any $\mathbf{p}_i$ and
$\bar{\mathbf{p}}_i\in\mathcal{P}_i$ and for fixed
$\mathbf{p}_{-i}$,
\begin{align}
R_i(\mathbf{p}_i,\mathbf{p}_{-i})-R_i(\bar{\mathbf{p}}_i,\mathbf{p}_{-i})
=P(\mathbf{p}_i,\mathbf{p}_{-i})-P(\bar{\mathbf{p}}_i,\mathbf{p}_{-i})\label{eqPotentialProperty2}.
\end{align}

 We call the function $P(\mathbf{p})$ the {\it
potential function} associated with the game $G$. Due to the
properties \eqref{eqPotentialProperty1} and
\eqref{eqPotentialProperty2}, we call the game $G$ a {\it potential
game}. From \cite{monderer96}, \cite{deb08} and \cite{scutari06}, we
have the following theorem.

\newtheorem{T1}{Theorem}
\begin{T1}\label{theoremPotential}
{\it A potential game, say
$\mathcal{G}=\{{\Omega},{\chi},\{U_i\}_{i\in\Omega}\}$, admits at
least one pure-strategy NE. If the potential function $P(.)$
associated with the potential game is concave, then a feasible
strategy $\mathbf{x}^*\in\chi$ is a NE of the game if and only if it
maximizes the potential function, i.e.,
$\mathbf{x}^*\in\arg_{\mathbf{x}\in{\chi}}\max P(\mathbf{x})$.}
\end{T1}

In light of the above theorem, we immediately have the following
Corollary.
\newtheorem{C1}{Corollary}
\begin{C1}\label{corPotential}
{\it $\mathbf{p}^*$ is a NE of the game $G$ if and only if
\begin{align}
\mathbf{p}^*\in\arg\max_{\mathbf{p}\in\mathcal{P}}
\frac{1}{K}\sum_{k=1}^{K}\left(\log\left(n(k)+\sum_{i=1}^{N}|h_i(k)|^2p_i(k)\right)-\log(n(k))\right).
\label{eqOptimizationPotential}
\end{align}}
\end{C1}

\subsection{NE as Optimum Input Strategies that Maximize the Network
Sum Rate}\label{subNEOPTInput}

It turns out that the potential function \eqref{eqPotential} has a
nice physical interpretation. We show in this subsection that it can
be related to the maximum sum rate achievable for the considered
network. In order to make the above statement precise, we digress a
little to consider the following sum capacity maximization problem
of the vector MAC system.

Consider a vector MAC communication system \cite{yu04} with $N$
users and a single AP. The users and the AP are both equipped with
$K$ antennas. Assume the available bandwidth is $\frac{1}{K}$. Let
$\widehat{\mathbf{x}}_k$ be user $k$'s transmitting vector signal;
let $\widehat{\mathbf{H}}_k$ be a $K\times K$ matrix that represents
the communication channel between user $k$ and the AP; let
$\widehat{\mathbf{y}}$ be the aggregated received signal at the AP;
let $\widehat{\mathbf{z}}$ be the additive Gaussian noise with
covariance matrix $\bfSigma_z$ . Then $\widehat{\mathbf{y}}$ can be
expressed as follows \footnote{Clearly the received signal
$\widehat{\mathbf{y}}$ has a similar form with that of the received
signal in considered single AP network (cf.
\eqref{eqMultiChannelMAC}), consequently, the results derived in
this vector MAC system are instrumental in analyzing our single AP
network. }:
\begin{align}
\widehat{\mathbf{y}}=\sum_{i\in\mathcal{N}}\widehat{\mathbf{H}}_i\widehat{\mathbf{x}}_i+
\widehat{\mathbf{z}}.\label{eqVectorMAC}
\end{align}
Let $\bfSigma_i\triangleq
E[\widehat{\mathbf{x}}_k\widehat{\mathbf{x}}^{\intercal}_k]$ denote
user $i$'s transmission/input covariance.  The users are constrained
in their individual power output, i.e., the input covariance of the
users should satisfy $tr(\bfSigma_i)\le \bar{p}_i,
~\forall~i\in\mathcal{N}$. The capacity-achieving input distribution
is known to be a complex Gaussian distribution, and the optimum set
of input covariances $\{\bfSigma^*_i\}_{i\in\mathcal{N}}$ that
maximize the capacity of this system can be found by solving the
following problem \cite{yu04}, \cite{cover05}:
\begin{align}
\max & ~~C(\bfSigma)\triangleq
\frac{1}{K}\left(\log\left|\sum_{i\in\mathcal{N}}\widehat{\mathbf{H}}_i\bfSigma_i\widehat{\mathbf{H}}^{\intercal}_i+\bfSigma_z
\right|-\log|\bfSigma_z|\right)& \label{eqOptimizationAP}\\
s.t. & ~~~tr(\bfSigma_i)\le \bar{p}_i~~~\forall~i\in\mathcal{N} \nonumber\\
  &~~~\bfSigma_i\succeq 0~~\forall~i\in\mathcal{N} \nonumber
\end{align}
where $\bfSigma\triangleq\{\bfSigma_i\}_{i\in\mathcal{N}}$ indicates
the joint transmission covariance matrix. Notice, that the objective
function $C(\bfSigma)$ is a concave function \cite{cover05}, hence,
it admits a unique maximum value in the feasible region. However,
this function is generally not strictly convex, and there are a
(connected) set of optimum points that achieve such optimum value.

We have the following proposition regarding the input covariance
matrices $\bfSigma^*$ that maximizes the sum rate of the system.

\newtheorem{P1}{Proposition}
\begin{P1}\label{propDiagonal}
{\it If $\widehat{\mathbf{H}}_i$ is diagonal for each $i$, and if
$\bfSigma_z$ is diagonal, then there must exist a set of diagonal
input covariance matrices $\{\bfSigma^*_i\}_{i\in\mathcal{N}}$ that
is the optimum solution of the problem \eqref{eqOptimizationAP}.}
\end{P1}
\begin{proof}
We prove this proposition by contradiction. Consider the following
convex problem:
\begin{align}
\max & ~~C(\bfSigma)\triangleq
\frac{1}{K}\left(\log\left|\sum_{i\in\mathcal{N}}\widehat{\mathbf{H}}_i\bfSigma_i\widehat{\mathbf{H}}^{\intercal}_i+\bfSigma_z
\right|-\log|\bfSigma_z|\right)& \label{eqOptimizationAPRestricted}\\
s.t. & ~~~tr(\bfSigma_i)\le \bar{p}_i,~~~\forall~i\in\mathcal{N} \nonumber\\
  &~~~\bfSigma_i\succeq 0,~~~\forall~i\in\mathcal{N} \nonumber\\
  &~~~\bfSigma_i \textrm{~is diaganol}, ~~~\forall~i\in\mathcal{N}. \nonumber
\end{align}
Suppose the set of diagonal matrices
$\{\bfSigma^*_i\}_{i\in\mathcal{N}}$ is an optimum solution of the
problem \eqref{eqOptimizationAPRestricted}, but it is not an optimum
solution of the problem \eqref{eqOptimizationAP}. For each user
$i\in\mathcal{N}$, let $\widetilde{\bfSigma}_i$ be the (unique)
solution of the following optimization problem:
\begin{align}
\max_{\bfSigma_i} &~~~
\frac{1}{K}\left(\log\left|\widehat{\mathbf{H}}_i\bfSigma_i\widehat{\mathbf{H}}^{\intercal}_i+\sum_{j\ne
i}\widehat{\mathbf{H}}_j\bfSigma^*_j\widehat{\mathbf{H}}^{\intercal}_j+\bfSigma_z
\right|-\log|\bfSigma_z| \right)\label{eqOptimizationAPSingleUser}\\
s.t. & ~~~tr(\bfSigma_i)\le \bar{p}_i \nonumber\\
  &~~~\bfSigma_i\succeq 0\nonumber.
\end{align}
Then from Theorem 1 of \cite{yu04}, and the assumption that
$\{\bfSigma^*_i\}_{i\in\mathcal{N}}$ is not optimum solution to the
problem \eqref{eqOptimizationAP}, there must be at least one user
${i}$ that, by changing its input covariance matrices from
$\bfSigma_i^*$ to $\widetilde{\bfSigma}_i$, it can strictly improve
the objective function, i.e., $\exists~i\in\mathcal{N}$, such that
$C(\widetilde{\bfSigma}_i,\bfSigma^*_{-i})>C(\bfSigma^*)$.

Let us now find the optimum solution $\widetilde{\bfSigma}_i$ of the
problem \eqref{eqOptimizationAPSingleUser}. Let
$\mathbf{N}\triangleq\sum_{j\ne
i}\widehat{\mathbf{H}}_j\bfSigma^*_j\widehat{\mathbf{H}}^{\intercal}_j+\bfSigma_z$,
then $\mathbf{N}$ is diagonal because $\bfSigma_z$,
$\{\bfSigma^*_j\}_{j\ne i}$ and $\{\widehat{\mathbf{H}}_i\}_{i\ne
j}$ are all diagonal. Clearly it is also semi-definite. Define a
matrix $\mathbf{Q}$ with its elements satisfying:
\begin{align}
[\mathbf{Q}]_{i,i}= \left\{ \begin{array}{ll}
\frac{1}{\sqrt{[\mathbf{N}]_{i,i}}} &\textrm{if } [\mathbf{N}]_{i,i}\ne 0\\
0  &\textrm{otherwise } \\
\end{array} \right..
\end{align}
Then solving problem \eqref{eqOptimizationAPSingleUser} is
equivalent to maximizing the function
$\log\left|\mathbf{Q}\widehat{\mathbf{H}}_i\bfSigma_i\widehat{\mathbf{H}}^{\intercal}_i\mathbf{Q}+\mathbf{I}
\right|$. From Hadamard's inequality \cite{cover05} and the fact
that $\mathbf{Q}\widehat{\mathbf{H}}_i$ is diagonal, we have that:
\begin{align}
\log\left|\mathbf{Q}\widehat{\mathbf{H}}_i\bfSigma_i\widehat{\mathbf{H}}^{\intercal}_i\mathbf{Q}+\mathbf{I}
\right|\le\sum_{k\in\mathcal{K}}\log\left(1+[\widehat{\mathbf{H}}_i]_{k,k}^2[\mathbf{Q}]^2_{k,k}[\bfSigma_i]_{k,k}\right)\nonumber
\end{align}
where the equality is achieved if and only if $\bfSigma_i$ is
diagonal, and satisfying $0\le\sum_{k=1}^{K}[\bfSigma_i]_{k,k}\le
\bar{p}_i$. Consequently, we conclude that the optimum covariance
matrix $\widetilde{\bfSigma}_i$ of the problem
\eqref{eqOptimizationAPSingleUser} is a diagonal matrix. However,
this contradicts the assumption that the set of covariance matrices
$\{\bfSigma^*_i\}_{i\in\mathcal{N}}$ is an optimum solution of the
problem \eqref{eqOptimizationAPRestricted}, because the new set of
covariance matrices $\{\bfSigma^*_{-i}, \widetilde{\bfSigma}_i\}$ is
a feasible solution to the problem
\eqref{eqOptimizationAPRestricted}, but
$C(\widetilde{\bfSigma}_i,\bfSigma^*_{-i})>C(\bfSigma^*)$.
\end{proof}

The proof of Proposition \ref{propDiagonal} points out that when
$\{\widehat{\mathbf{H}}_i\}_{i\in\mathcal{N}}$ and $\bfSigma_z$ are
diagonal, any solution to the optimization problem
\eqref{eqOptimizationAPRestricted} must be an optimal solution to
the original problem \eqref{eqOptimizationAP}. In this case, the
users only need to select the amount of power on each antenna (power
loading transmission scheme) to achieve the maximum sum rate, and
the objective function $C(\bfSigma)$ of
\eqref{eqOptimizationAPRestricted} can be reduced to:{\small
\begin{align}
&\frac{1}{K}\log\left|\sum_{i\in\mathcal{N}}\widehat{\mathbf{H}}_i\bfSigma_i\widehat{\mathbf{H}}^{\intercal}_i+\bfSigma_z
\right|-\frac{1}{K}\log|\bfSigma_z| \nonumber\\
&=\frac{1}{K}\sum_{k=1}^{K}\left(\log\left([\bfSigma_z]_{k,k}+\sum_{i=1}^{N}
\left|[\widehat{\mathbf{H}}_i]_{k,k}\right|^2[\bfSigma_i]_{k,k}\right)-\log[\bfSigma_z]_{k,k}\right).\nonumber
\end{align}}
Clearly, when the channel matrices and the noise matrix are all
diagonal, and the users choose the diagonal transmission strategy,
the $N$-user $K$-antenna vector MAC channel introduced above is
equivalent to the our previously considered $N$-CU $K$-channel
single AP network (as can be seen from the equivalence of equation
\eqref{eqMultiChannelMAC} and \eqref{eqVectorMAC}), with the
following correspondence of parameters: $
\bfSigma_i=diag([p_i(1),\cdots,p_i(K)])~~\forall~ i\in\mathcal{N}$,
$
\widehat{\mathbf{H}}_i=diag([h_i(1),\cdots,h_i(K)])~~\forall~i\in\mathcal{N}$
and $\bfSigma_z=diag([n(1),\cdots, n(K)])$.

The vector MAC capacity maximization problems
\eqref{eqOptimizationAP} or \eqref{eqOptimizationAPRestricted} can
be readily seen as equivalent to the potential function maximization
problem \eqref{eqOptimizationPotential}, and the maximum value of
the potential function $P(\mathbf{p})$, say $P^*$, corresponds to
the maximum sum rate achievable for the considered N-CU, K-channel,
single AP network. Corollary \ref{corPotential} implies that any NE
of the game $G$ maximizes the potential function $P(\mathbf{p})$
among the feasible solutions. Consequently it is also an optimal
solution of \eqref{eqOptimizationAPRestricted}, thus can be viewed
as a set of optimal input covariance matrices (or optimum power
loading scheme because of the diagnonality) that maximizes the
achievable sum rate of the system.

However, this result does not imply that the sum
of the CUs' rate at a NE achieves the optimal system sum rate. 
%
We should point out here that in general, one needs to have both
{\it optimal transmission strategy} and {\it optimal receiving
strategy} to be able to achieve the MAC capacity. In our context,
this is to say that in general, at a NE of the game $G$, although
the CUs' transmission strategy is optimal, the sum of individual
CUs' rate should be less than the maximum achievable sum rate of the
system (notice, that in our considered network, the assumption is
that only single-user receiver is implemented at the AP, which is
obviously not an optimal receiving strategy).

However, we observe that if a NE of the game $G$ represents a FDMA
strategy, then the maximum system sum rate is achieved using only
single-user receiver at the AP. More specifically, if a NE
$\mathbf{p}^*$ represents the FDMA strategy, then there is at most a
single user $i\in\mathcal{N}$ transmitting on each channel $k$
(i.e., with $\mathbf{p}^*_{i}(k)>0$). Define the index set
$\mathcal{I}(k)\triangleq\{i: {p}^*_i(k)>0\}$, we have that under a
FDMA transmission strategy $|\mathcal{I}(k)|\le
1,~~\forall~k\in\mathcal{K}$. Clearly in this case, the sum rate of
the users (when the AP uses single-user receiver) achieves the
maximum sum rate of the network:
\begin{align}
&\sum_{i\in\mathcal{N}}R_i(\mathbf{p}^*_i,\mathbf{p}^*_{-i})=\sum_{i\in\mathcal{N}}\sum_{k=1}^{K}\log\left(1+\frac{{p}^*_i(k)|h_i(k)|^2}{n(k)+\sum_{j\ne
i}{p}_j(k)|h_j(k)|^2}\right)\nonumber\\
&\stackrel{(a)}=\sum_{k=1}^{K}\sum_{i\in\mathcal{I}(k)}
\log\left(1+\frac{{p}^*_{i}(k)|h_{i}(k)|^2}{n(k)}\right)\nonumber\\
& \stackrel{(b)}=
\sum_{k=1}^{K}\left(\log\left(n(k)+\sum_{i=1}^{N}|h_i(k)|^2p^*_i(k)\right)-\log(n(k))\right)=P(\mathbf{p}^*)
\end{align}
where both $(a)$ and $(b)$ are from the FDMA property of the NE
$\mathbf{p}^*$.

The question remains as in what situation does the NE of the game
$G$ represents the FDMA transmission strategy. In the next
subsection, we provide an answer to this question by looking at the
situation when the available spectrum is arbirarily finely divided.
i.e., $K\gg N$.

\subsection{The Asymptotic Optimality of the
NE}\label{subNEAsymptotic} In this subsection, we consider the
asymptotic situation in which the available bandwidth is arbitrarily
finely divided. In this case, the transmission rate for each CU $i$
can be expressed as \cite{cover05}:
\begin{align}
R_i(\mathbf{p}_i,\mathbf{p}_{-i})=\int_{0}^{1}\log\left(1+\frac{{p}_i(w)|h_i(w)|^2}{n(w)+\sum_{j\ne
i}{p}_j(w)|h_j(w)|^2}\right)d w
\end{align}
where the channel gain $|h_i(w)|^2$ can be viewed as the channel
transfer function for CU $i$ to the AP; $n(w)$ becomes the spectral
density of the Gaussian noise experienced at the AP on channel $k$;
$\mathbf{p}_i$ denotes the transmit {\it power spectral density} of
CU $i$, i.e., $\mathbf{p}_i(w)$ indicates the amount of power CU $i$
transmits on frequency $w$. In this case, the sum power constraint
of each CU should be expressed as: $\int_0^1 {p}_i(w)d w\le
\bar{p}_i$, and the feasible region $\mathcal{P}_i$ becomes:
$\mathcal{P}_i\triangleq\{\mathbf{p}_i:\int_0^1 {p}_i(w)d w\le
\bar{p}_i, {p}_i(w)\ge 0~\forall~w\}$.

As before, a selfish CU $i$ is interested in solving the following
optimization problem:
\begin{align}
&\max_{\mathbf{p}_i\in\mathcal{P}_i}~\int_{0}^{1}\log\left(1+\frac{{p}_i(w)|h_i(w)|^2}{n(w)+\sum_{j\ne
i}{p}_j(w)|h_j(w)|^2}\right) d w\label{eqUserProblemContinuous}.
\end{align}

From the definition of the NE and the solution to the CU's utility
maximization problem \eqref{eqUserProblemContinuous}, individual
equilibrium transmit spectral $\mathbf{p}^*_i$ must satisfy:
\begin{align}
{p}^*_{i}(w)=\left[\sigma_i-\frac{n(w)+\sum_{j\ne
i}{p}^*_j(w)|h_j(w)|^2}{|h_i(w)|^2}\right]^{+},~\forall~w.
\end{align}

We have the following theorem characterizing the system equilibrium
transmit spectral $\{\mathbf{p}^*_i\}_{i\in\mathcal{N}}$.
\newtheorem{T3}{Theorem}
\begin{T1}\label{theoremAsymptotic}
{\it When the available spectrum is arbitrarily finely divided, and
the channel gains are generated according to some continuous
distribution, then any NE of the game $G$ represents a FDMA
transmission strategy (with probability 1). Moreover, any such NE is
efficient, in the sense that the sum of individual users' rates
achieves the maximum system sum rate.}
\end{T1}
\begin{proof}
We first show, by contradiction, that any NE represents the FDMA
strategy.

Suppose for some channel realization
$\{|h_i(w)|^2\}_{i\in\mathcal{N}}$, in the NE of the game a set of
CUs $\mathcal{M}\subseteq\mathcal{N}$ are using the frequency $w$.
In another words, we assume the following:
\begin{align}
{p}^*_{m}(w)=\sigma_m-\frac{n(w)+\sum_{j\ne
m,j\in\mathcal{M}}{p}^*_j(w)|h_j(w)|^2}{|h_m(w)|^2}>0,~\forall~m\in\mathcal{M}.
\end{align}
Then the following is true for all $m\in\mathcal{M}$:
\begin{align}
n(w)+\sum_{j\in\mathcal{M}}{p}^*_j(w)|h_j(w)|^2=\sigma_m
|h_m(w)|^2.\label{eqSharingContradiction}
\end{align}
Thus, for an arbitrary pair of CUs $m_1,~m_2\in\mathcal{M}$:
$\sigma_{m_1}|h_{m_1}(w)|^2=\sigma_{m_2}|h_{m_2}(w)|^2$. However,
this equality is satisfied with probability zero (see the proof of
Theorem 1 of \cite{lai08}), because of the fact that $\sigma_{m_1}$
and $\sigma_{m_2}$ are constants, and that the channel coefficients
are random variables drawn from continuous distributions (Rayleigh
distribution or Rician distribution in fading channels). In summary,
we claim that the equilibrium transmit power spectral
$\{\mathbf{p}_i^*\}$ follows a FDMA scheme with probability 1.

Due to the above FDMA frequency allocation scheme, when
${p}^*_i(w)>0$, it must be true that:
\begin{align}
{p}^*_i(w)&=\left[\sigma_i-\frac{n(w)}{|h_i(w)|^2}\right]^+=
\sigma_i-\frac{n(w)}{|h_i(w)|^2}\label{eqEquilibriumCondition1}.
\end{align}
Consequently, we can have, for $j\ne i$ (thus ${p}^*_j(w)=0$):
\begin{align}
&{p}^*_j(w)=\left[\sigma_j-\frac{n(w)+|h_i(w)|^2p^*_i(w)}{|h_j(w)|^2}\right]^+=0\nonumber\\
&\Longrightarrow
\sigma_j-\frac{n(w)+|h_i(w)|^2{p}_i^*(w)}{|h_j(w)|^2}\le 0
\stackrel{(a)}\Longrightarrow \sigma_j |h_j(w)|^2\le\sigma_i
|h_i(w)|^2 \label{eqEquilibriumCondition2}
\end{align}
where $(a)$ is because of \eqref{eqEquilibriumCondition1}. From
Theorem 2 of \cite{cheng93}, we know that the $N$-user FDMA scheme
maximizes the system sum-rate for a Gaussian multiple access channel
if the following is true:
\begin{align}
&\int_0^1 {p}_i(w) d w=\bar{p}_i,~\forall~i\in\mathcal{N}\label{eqOptimalityCondition1}\\
&{p}_i(w)=\left\{ \begin{array}{ll}
\left[b_i-\frac{n(w)}{|h_i(w)|^2}\right]^+,&\textrm{if~}b_i
|h_i(w)|^2\ge b_j |h_j(w)|^2,\forall~j\ne i\\
0,&\textrm{otherwise.}\\
\end{array} \right.\label{eqOptimalityCondition2}
\end{align}

Comparing
\eqref{eqEquilibriumCondition1}--\eqref{eqEquilibriumCondition2}
with \eqref{eqOptimalityCondition1}--\eqref{eqOptimalityCondition2},
we can readily identify that $\sigma_i=b_i$, and we conclude that
$\{\mathbf{p}^*_i\}_{i\in\mathcal{N}}$ achieves the maximum system
sum rate.
\end{proof}

We see from the above derivation that the optimum channel assignment
should take into consideration the following three factors
\cite{yu02a}: 1) users' channel quality; 2) users' power budget; 3)
the noise power. The results derived in Theorem
\ref{theoremAsymptotic} is desirable because in practical
multi-carrier systems (e.g. OFDM system), the number of channels is
indeed very large compared with the number of users in the system.
Consequently, the NE of the spectrum sharing game $G$ represents a
desirable outcome in which the CUs in the network share the spectrum
efficiently. Interestingly, the authors of \cite{lai08} has shown
that the NE for a uplink power control game represents a
time-sharing strategy (which can be viewed as dual to our FDMA
strategy), but in a 2-user fading channel system which is very
different from the system we consider.

Now the question becomes how such equilibrium point(s) can be
reached by individual CU in a distributed fashion. In the next
section, we provide three algorithms for such purpose.


\section{The Proposed Algorithms and
Convergence}\label{secAlgorithm}


From the argument in the previous section, we see that finding the
NE of the game $G$ is equivalent to finding
$\mathbf{p}^*\in\arg\max_{\mathbf{p}\in\mathcal{P}}\max
P(\mathbf{p})$. This is a convex problem and can be solved in a
centralized way if all the parameters of the system (e.g.,
$\left\{|h_i(k)|^2\right\}$, $\left\{\bar{p}_i\right\}$) are known.
However, in a distributed environment, where the CUs are selfish,
uncoordinated and not well informed of other CUs' channel
coefficients and power budgets, it is not immediately clear how to
find such NE point in a distributed fashion.

\subsection{Inapplicability of Conventional IWF
Algorithm}\label{subInapplicableIWF}

We first notice that our model of the network is a special case of a
more general network with Gaussian interference channel that has
been extensively studied recently, for example, in
\cite{scutari08a}, \cite{luo06b}, \cite{scutari08b},
 \cite{shum07}. In those works, the CUs are
transmitter-receiver pairs, and they are interested in allocating
their limited transmission power on the set of channels
$\mathcal{K}$ to maximize their individual transmission rate. We
refer to this network as a Peer-to-Peer (PP) network, while
referring to our network as a Access Point (AP) network. In the PP
network, we use $|H_{i,j}(k)|^2$ to denote the channel gain from CU
$i$'s transmitter to CU $j$'s receiver on the $k^{th}$ channel; we
use $n_i(k)$ to denote the environmental noise power at the receiver
of CU $i$ on channel $k$. An individual CU $i$, by transmiting
$p_i(k)$ on the $k^{th}$ channel, contributes to every other CU
$j\ne i$ in the network the amount of $|H_{i,j}(k)|^2p_i(k)$
interference at their respective receivers.

Now consider the scenario where all the CUs' receivers are
co-located. In this case, for a particular CU $i$, the set of
channel coefficients $\{|H_{i,j}(k)|^2\}_{j\ne i}$ become equal to
the value of $|H_{i,i}(k)|^2$; the set of environment noises
$\{n_i(k)\}_{i\in\mathcal{N}}$ can be considered equal because the
receivers are located at the same place. Consequently, the PP
network is equivalent to the AP network.

At this point, we might come to the conclusion that the distributed
algorithms developed for the PP network automatically works in the
AP network, after all, the PP case is more general than the AP case.
However, we show in the following that this is not true. As a matter
of fact, the sufficient conditions for the convergence of most
algorithms proposed for the PP network are not satisfied in the AP
network. As an example, we consider the sufficient condition for the
simultaneous IWF algorithm proposed in \cite{scutari08b}.

%
%

Define $K$ nonnegative matrices $\mathbf{H}(k)\in
\mathbb{R}^{N\times N}_{+}$ with their elements defined as follows:
\begin{align}
[\mathbf{H}]_{q,r}(k)\triangleq \left\{ \begin{array}{ll}
\frac{|H_{r,q}(k)|^2}{|H_{q,q}(k)|^2}&\textrm{if } r\ne q\\
0 &\textrm{otherwise. } \\
\end{array} \right. \label{eqDefineH}
\end{align}
Define another nonnegative matrix $\mathbf{H}^{\max}\in
\mathbb{R}^{N\times N}_{+}$ as follows:
\begin{align}
[\mathbf{H}]^{\max}_{q,r}\triangleq \left\{ \begin{array}{ll}
\max_k\{\frac{|H_{r,q}(k)|^2}{|H_{q,q}(k)|^2}\}&\textrm{if } r\ne q\\
0 &\textrm{otherwise. } \\
\end{array} \right. \label{eqDefineH}
\end{align}

From Theorem 1 in \cite{scutari08b}, we have that the simultaneous
IWFA algorithm converges to the unique NE of the game if the
following is true: $\rho(\mathbf{H}^{\max})<1
\label{eqHRadiusCondition}$. In the following, we prove that in AP
scenario, this condition  can not be satisfied.

From the Perron-Frobenius Theorem \cite{bertsekas97}, we have that
there must exist a $N\times1$ vector $\mathbf{w}>0$, such that
$||\mathbf{H}^{\max}||^{\mathbf{w}}_{\infty}=\rho(\mathbf{H}^{\max})$,
where $||\mathbf{A}||^\mathbf{w}_{\infty}$ is the maximum norm of a
matrix $\mathbf{A}$, and is defined as follows:
\begin{align}
||\mathbf{A}||^\mathbf{w}_{\infty}\triangleq
\max_{q}\frac{1}{w_q}\sum_{r=1}^{Q}[\mathbf{A}]_{q,r}w_r,~\mathbf{A}\in\mathbf{R}^{Q\times
Q}.
\end{align}

We next show that in the AP case, there could be no positive vector
$\mathbf{w}$ satisfying
$||\mathbf{H}^{\max}||^{\mathbf{w}}_{\infty}<1$. Note that we have
for all $k\in\mathcal{K}$, $0\le \mathbf{H}(k)\le\mathbf{H}^{\max}$
componentwise, which implies
$||\mathbf{H}(k)||^{\mathbf{w}}_{\infty}\le||\mathbf{H}^{\max}||^{\mathbf{w}}_{\infty}$
(\cite{bertsekas97}, Chapter 2, Proposition 6.2). Consequently, it
is sufficient to prove that there exists $k\in\mathcal{K}$, such
that for all $\mathbf{w}>0$, we must have
$||\mathbf{H}(k)||^{\mathbf{w}}_{\infty}\ge 1$.

Choose any $k\in\mathcal{K}$. Suppose there exists $\mathbf{w}>0$
such that $||\mathbf{H}(k)||^{\mathbf{w}}_{\infty}< 1$. This implies
that: $ \max_{j}\frac{1}{w_j}\sum_{i\ne j}
\frac{|H_{i,j}(k)|^2}{|H_{j,j}(k)|^2} w_i <1$. Then it must be true
that for every $j\in\mathcal{N}$, $ \frac{1}{w_j}\sum_{i\ne j}
\frac{|H_{i,j}(k)|^2}{|H_{j,j}(k)|^2} w_i <1 $, which is equivalent
to say that the following $N$ inequalities are true simultaneously:
\begin{align}
\sum_{i\ne j} |H_{i,j}(k)|^2 w_i
<{|H_{j,j}(k)|^2}w_j,~\forall~j\in\mathcal{N}.\label{eqHContradiction}
\end{align}

Recall that when reduced to AP configuration, we have that for all
$j\ne i$, $|H_{i,j}(k)|^2=|H_{i,i}(k)|^2$. Using this equality and
adding up $N$ inequalities in \eqref{eqHContradiction}, we must
have:
\begin{align}
(N-1)\sum_{i\in\mathcal{N}} |H_{i,i}(k)|^2
w_i<\sum_{j\in\mathcal{N}} |H_{j,j}(k)|^2 w_j.
\end{align}
Because all the channel coefficients are greater than $0$, the above
inequality can not be satisfied for any $\mathbf{w}>\mathbf{0}$.
Consequently, we prove that there does not exist any
$\mathbf{w}>\mathbf{0}$ such that
$||\mathbf{H}(k)||^{\mathbf{w}}_{\infty}< 1$. Thus, we must have
that $||\mathbf{H}(k)||^{\mathbf{w}}_{\infty}\ge 1$, which in turn
says that for all $\mathbf{w}>\mathbf{0}$, we must have
$||\mathbf{H}^{\max}||^{\mathbf{w}}_{\infty}\ge 1$, and this implies
$\rho(\mathbf{H}^{\max})\ge 1$. We note further that since
$\rho(\mathbf{H}^{\max})\le||\mathbf{H}^{\max}||$ for any norm
(Prop.A.20 in \cite{bertsekas97}), we must have that
$||\mathbf{H}^{\max}||\ge 1$ for arbitrary norm. Moreover, we can
show similarly that for arbitrary norm, $||\mathbf{H}(k)||\ge
1,~\forall~k$, and thus $\rho(\mathbf{H}(k))\ge 1~\forall~k$.

In order to further explain the reason why, in general, algorithms
for the PP configuration fail to work in our AP configuration, we
observe that almost all the algorithms designed for PP configuration
rely on some restrictive conditions of the channel gains to ensure
the {\it uniqueness} of the equilibrium. For example, in
\cite{scutari08a}, the condition
$\rho(\mathbf{H}(k))<1,~\forall~k\in\mathcal{K}$ ensures the NE of
the power allocation game is unique. However as we see in our
previous argument, in the AP configuration such condition is not
true {\it for any realization of the channel gains}. As a matter of
fact, a straightforward consequence of Corollary \ref{corPotential}
is that in general the AP configuration admits {\it a (connected)
set} of equilibrium solutions, as the objective function of the
optimization problem \eqref{eqOptimizationPotential} is concave, but
not strictly concave. A simple example illustrates this point.

\newtheorem{E1}{Example}
\begin{E1}\label{exMultipleEquilibrium}
{\it Consider the network with $N=2$ CUs, $K=2$ channels. Let
$|h_1(1)|^2=|h_2(1)|^2=1$, $|h_1(2)|^2=|h_2(2)|^2=2$, $n(1)=n(2)=1$,
and let $\bar{p}_1=\bar{p}_2=1$. We can show that both the following
two system power profiles $\widetilde{\mathbf{p}}$ and
$\widehat{\mathbf{p}}$ are the NE for the game related to this
network:
\begin{align}
\widetilde{p}_1(1)=\frac{3}{4},~\widetilde{p}_1(2)=\frac{1}{4};~\widetilde{p}_2(1)=0,~\widetilde{p}_1(2)=1;
\end{align}
and
\begin{align}
\widehat{p}_1(1)=0,~\widehat{p}_1(2)=1;~\widehat{p}_2(1)=\frac{3}{4},~\widehat{p}_1(2)=\frac{1}{4}.
\end{align}
Clearly, from the concavity of the potential function, all the
convex combinations of the solutions $\widetilde{\mathbf{p}}$ and
$\widehat{\mathbf{p}}$ also maximize the potential function, hence
they are also NEs of the game $G$. }
\end{E1}

We conclude the above argument by saying that although the AP
network indeed is a special case of the more general PP network, for
which distributed algorithms have been developed to reach the NE,
these algorithms may not be directly applicable to the AP scenario.
Indeed, we will see later in the simulation section, that by
applying the simultaneous IWF algorithm directly to the AP network
results in divergence.

\subsection{Proposed Algorithm based on IWF: Weak Convergence}
We now proceed to develop algorithms so that the CUs in the single
AP network can distributedly reach the NE. In the following we
propose two such algorithms.

{\bf Algorithm 1: Averaged Iterative-Water Filling Algorithm (A-IWF)}:\\
 In each iteration $t$, the CUs do the
following.\\
1) Calculate the best reply power allocation: {
\begin{align}
\Phi^k_i(\mathbf{p}^t_{-i}) &\triangleq
\left[\frac{1}{\sigma_i}-\frac{n_i(k)+\sum_{j\ne
i}|h_j(k)|^2p^t_j(k)}{|h_{i}(k)|^2}\right]^+,
\forall~k\in\mathcal{K}
\end{align}
where $\sigma_i$ ensures
$\sum_{k\in\mathcal{K}_w}{\Phi^k_i({\mathbf{p}}^t_{-i})}=\bar{p}_i$,
and let
$\bfPhi_i(\mathbf{p}^t_{-i})\triangleq\left[\Phi^1_i({\mathbf{p}}^t_{-i}),\cdots, \Phi^K_i({\mathbf{p}}^t_{-i})\right]^{\intercal}$.\\
2) Adjust their power profiles simultaneously according to:
\begin{align}
\mathbf{p}^{t+1}_{i}&=(1-\alpha_t)\mathbf{p}^{t}_{i}+\alpha_t\bfPhi_i(\mathbf{p}^t_{-i})
\end{align}}
where the sequence $\{\alpha_t\}_{t=1}^{\infty}$ satisfy
$\alpha_t\in(0,1)$ and :
\begin{align}
\lim_{T\to\infty}\sum_{t=1}^{T}\alpha_t=\infty,
~\lim_{T\to\infty}\sum_{t=1}^{T}\alpha^2_t<\infty.\label{eqAlphaProperty}
\end{align}

{\bf Algorithm 2: Sequential Iterative-Water Filling Algorithm (S-IWF)}:\\
{\ In each iteration $t$, the CUs adjust their power profiles
sequentially\footnote{By ``sequential" we mean that the CUs in the
set $\mathcal{N}$ take turns in changing their power allocation, and
only a single CU gets to act at time $t$. All other CUs $j\ne i,
j\in\mathcal{N}$ keep their power allocation as in time $t-1$.}
according to:
\begin{align}
\mathbf{p}^{t+1}_{i}&=\bfPhi_i(\mathbf{p}^t_{-i}).
\end{align}
}
The convergence properties of the above two algorithms are stated
in the following two propositions.

\newtheorem{P2}{Proposition}
\begin{P1}\label{propAIWF}
{\it If all the CUs in the network employ A-IWF algorithm, then
their individual power profiles converge to the set of NE of game
$G$.}
\end{P1}
\begin{proof}
Define
$\bfPhi(\mathbf{p})\triangleq[\bfPhi_1(\mathbf{p}_{-1}),\cdots,\bfPhi_N(\mathbf{p}_{-N})]^{\intercal}$.
Define $\mathbf{s}(\mathbf{p})=\bfPhi(\mathbf{p})-\mathbf{p}$. Then
from the system point of view the A-IWF algorithm can be written
concisely as:
\begin{align}
\mathbf{p}^{t+1}=(1-\alpha_t)\mathbf{p}^t+\alpha_t\bfPhi(\mathbf{p}^t)=\mathbf{p}^t+\alpha_t\mathbf{s}(\mathbf{p}^t).
\end{align}

We first introduce two lemmas. The proof of Lemma \ref{lemmaAscend}
can be found in Appendix \ref{appTheoremAIWF}, and we omit the proof
of Lemma \ref{lemmaLip} for brevity.

\newtheorem{L1}{Lemma}
\begin{L1}\label{lemmaAscend}
{\it There must exist a constant $M$, with $0<M<\infty$, such that $
\mathbf{s}(\mathbf{p})^{\intercal}\triangledown_{\mathbf{p}}P(\mathbf{p})\ge
M ||\mathbf{s}(\mathbf{p})||^2$.\label{eqM}}
\end{L1}

\newtheorem{L2}{Lemma}
\begin{L1}\label{lemmaLip}
{\it For two arbitrary vectors $\mathbf{p}\in\mathcal{P}$ and
$\bar{\mathbf{p}}\in\mathcal{P}$, and for arbitrary norm $||.||$,
there must exist two constants $0<D<\infty$, and $0<K<\infty$ such
that
\begin{align}
||\mathbf{s}(\mathbf{p})-\mathbf{s}(\bar{\mathbf{p}})||&\le
D||\mathbf{p}-\bar{\mathbf{p}}||\nonumber\\
||\triangledown P(\mathbf{p})-\triangledown
P(\bar{\mathbf{p}})||&\le K||\mathbf{p}-\bar{\mathbf{p}}||.
\end{align}}
\end{L1}

In order to conform to the convention in convex optimization, we
define the function $F(\mathbf{p})=-P(\mathbf{p})$, and we see that
$F(\mathbf{p})$ is convex.

Then from the well known Descent Lemma (Lemma 2.1 in
\cite{bertsekas97}), and Lemma \ref{lemmaLip} we have that:
\begin{align}
\hspace{-0.3cm}F(\mathbf{p}^{t+1})&\le
F(\mathbf{p}^t)+\alpha_t\mathbf{s}(\mathbf{p}^t)^{\intercal}\triangledown
F(\mathbf{p}^t)+\frac{K}{2}\alpha^2_t||\mathbf{s}(\mathbf{p}_t)||^2\label{eqTylor}\\
&\le
F(\mathbf{p}^t)-\alpha_t M||\mathbf{s}(\mathbf{p}_t)||^2+\frac{K}{2}\alpha^2_t||\mathbf{s}(\mathbf{p}_t)||^2\nonumber\\
&=F(\mathbf{p}^t)-\alpha_t ||\mathbf{s}(\mathbf{p}_t)||^2(M-\alpha_t
\frac{K}{2}).
\end{align}
Because $\alpha_t$ goes to $0$, then when $t$ large enough,
$M-\alpha_t\frac{K}{2}>0$, and $F(\mathbf{p}^t)$ is monotonically
decreasing. Combined with the fact that $F(\mathbf{p}^t)$ is lower
bounded, then $\{F(\mathbf{p}^t)\}_{t=1}^{\infty}$ is a convergent
sequence.

From \eqref{eqTylor}, we have that
\begin{align}
F(\mathbf{p}^{T+1})&\le
F(\mathbf{p}^0)+\sum_{t=0}^{T}\alpha_t\mathbf{s}(\mathbf{p}^t)^{\intercal}\triangledown
F(\mathbf{p}^t)+\sum_{t=0}^{T}\frac{K}{2}\alpha^2_t||\mathbf{s}(\mathbf{p}_t)||^2.\nonumber
\end{align}
It is clear that  $||\mathbf{s}(\mathbf{p}_t)||^2$ is upper bounded,
and we have $\lim_{T\to\infty}\sum_{t=1}^{T}\alpha^2_t<\infty$, so
$\lim_{T\to\infty}\sum_{t=0}^{T}\frac{K}{2}\alpha^2_t||\mathbf{s}(\mathbf{p}_t)||^2<\infty$.
Because $\lim_{T\to\infty}F(\mathbf{p}^{T+1})$ converges, we must
have
\begin{align}
\lim_{T\to\infty}\sum_{t=0}^{T}\alpha_t\mathbf{s}(\mathbf{p}^t)^{\intercal}\triangledown
F(\mathbf{p}^t)&>-\infty \nonumber\\
\lim_{T\to\infty}\sum_{t=0}^{T}\alpha_t\mathbf{s}(\mathbf{p}^t)^{\intercal}\triangledown
P(\mathbf{p}^t)&<\infty.\label{eqSummability}
\end{align}
From Lemma \ref{lemmaAscend}, we have
\begin{align}
M\lim_{T\to\infty}\sum_{t=0}^{T}\alpha_t||\mathbf{s}(\mathbf{p}^t)||^2\le\lim_{T\to\infty}\sum_{t=0}^{T}\alpha_t\mathbf{s}(\mathbf{p}^t)^{\intercal}\triangledown
P(\mathbf{p}^t)<\infty.\nonumber
\end{align}
Consequently it is clear that we must have
$\lim\inf_{t\to\infty}||\mathbf{s}(\mathbf{p}^t)||=0$. We show in
the following that in fact we have a stronger result that
$\lim_{t\to\infty}||\mathbf{s}(\mathbf{p}^t)||=0$. Suppose not, then
$\lim\sup_{t\to\infty}||\mathbf{s}(\mathbf{p}^t)||>0$. In this case
there must exist a $\epsilon>0$ such that the subsequences
$\{\tau(n): ||\mathbf{s}(\mathbf{p}^{\tau(n)})||<\epsilon,
||\mathbf{s}(\mathbf{p}^{\tau(n)+1})||\ge\epsilon\}$ and $\{u(n):
\epsilon\le ||\mathbf{s}(\mathbf{p}^{t})||\le 2\epsilon,
\forall~t\in(\tau(n), u(n)-1),~ ||\mathbf{s}(\mathbf{p}^{u(n)})||>
2\epsilon\}$ are both infinite.

For a specific $n$, the following is true:
\begin{align}
||\mathbf{s}(\mathbf{p}^{\tau(n)+1})||-||\mathbf{s}(\mathbf{p}^{\tau(n)})||&\le
D||\mathbf{p}^{\tau(n)+1}-\mathbf{p}^{\tau(n)}||\nonumber\\
&\le D\alpha_{\tau(n)}||\mathbf{s}(\mathbf{p}^{\tau(n)})||.
\end{align}
Thus, there exists a $N^*$ such that for all $n>N^*$, we must have
$||\mathbf{s}(\mathbf{p}^{\tau(n)})||\ge\frac{\epsilon}{2}$.

We also have the following:
\begin{align}
\epsilon&<
||\mathbf{s}(\mathbf{p}^{u(n)})||-||\mathbf{s}(\mathbf{p}^{\tau(n)})||
\le D||\mathbf{p}^{u(n)}-\mathbf{p}^{\tau(n)}||\nonumber\\
&\le
D\sum_{t=\tau(n)}^{t=u(n)-1}\alpha_t||\mathbf{s}(\mathbf{p}^{t})||\le
D\sum_{t=\tau(n)}^{t=u(n)-1}\alpha_t2 \epsilon
\end{align}
which implies
\begin{align}
\frac{1}{2D}<
\sum_{t=\tau(n)}^{t=u(n)-1}\alpha_t.\label{eqContradictionD}
\end{align}
From our previous derivation, we also have $
\lim_{T\to\infty}\sum_{t=0}^{T}\alpha_t||\mathbf{s}(\mathbf{p}^t)||^2<\infty.
$ Then for any $\delta>0$ there must exists a $\widehat{N}(\delta)$
such that for all $n>\widehat{N}(\delta)$: $
\sum_{t=\tau(n)}^{t=u(n)-1}\alpha_t||\mathbf{s}(\mathbf{p}^{t})||^2\le
\delta. $

Take $\delta=\frac{\epsilon^2 }{8 D}$, and take $n>\max\left\{{N^*,
\widehat{N}(\frac{\epsilon^2 }{8 D})}\right\}$, then we have
\begin{align}
\frac{\epsilon^2}{4}\sum_{t=\tau(n)}^{t=u(n)-1}\alpha_t\le\sum_{t=\tau(n)}^{t=u(n)-1}\alpha_t||\mathbf{s}(\mathbf{p}^{t})||^2
\le\frac{\epsilon^2 }{8 D}
\end{align}
which implies $\sum_{t=\tau(n)}^{t=u(n)-1}\alpha_t\le \frac{1}{2
D}$. This is a contradiction to \eqref{eqContradictionD}. Thus, we
conclude that $\lim\sup_{t\to\infty}||\mathbf{s}(\mathbf{p}^t)||=0$,
and consequently $\lim_{t\to\infty}||\mathbf{s}(\mathbf{p}^t)||=0$.

From $\lim_{t\to\infty}||\mathbf{s}(\mathbf{p}^t)||=0$ we see that
the limit point $\mathbf{p}^*_m$ of any converging subsequence of
$\{\mathbf{p}^t\}$, say $\{\mathbf{p}^{t_m}\}_{m=1}^{\infty}$, must
satisfy $\bfPhi(\mathbf{p}_m^*)=\mathbf{p}_m^*$, which is sufficient
condition to ensure that $\mathbf{p}_m^*$ is a NE of the game $G$.
Consequently $\mathbf{p}_m^*$ must maximize the function
$P(\mathbf{p})$ (from Corollary \ref{corPotential}), and this
implies that the entire sequence
$\{P(\mathbf{p}^t)\}_{t=1}^{\infty}$ converges to the value
$P^*\triangleq\max_{\mathbf{p}\in\mathcal{P}}P(\mathbf{p})$. It also
implies that the sequence $\{\mathbf{p}^t\}_{t=1}^{\infty}$ converge
the set of NE of the game $G$, or in other words, every limit point
of $\{\mathbf{p}^t\}_{t=1}^{\infty}$ is a NE of the game $G$.
\end{proof}

\newtheorem{P3}{Proposition}
\begin{P1}\label{propSIWF}
{\it If all the CUs in the network employ S-IWF algorithm, then
their individual power profiles converge to the set of NE of game
$G$. Moreover, the potential function
$\{P(\mathbf{p}^t)\}_{t=1}^{\infty}$ is non-decreasing with respect
to iteration step $t$, i.e., $P(\mathbf{p}^{t+1})\ge
P(\mathbf{p}^{t})$.}
\end{P1}
\begin{proof}
It is easy to see that this algorithm corresponds to the nonlinear
Gauss-Seidel algorithm in solving constrained optimization problem
\cite{bertsekas97}, thus the general theory for the convergence of
algorithm can be applied (e.g., \cite{bertsekas97} Proposition 3.9).
\end{proof}

The S-IWF algorithm is actually a simplification of the algorithm
proposed in \cite{yu04}. We introduce this algorithm here and
briefly discuss its convergence analysis because it will be useful
in our analysis in the second part of this paper. We need to point
out here that the convergence behaviors characterized for A-IWF and
S-IWF are {\it set convergence}, i.e., the distance between the
sequence $\{\mathbf{p}^t\}_{t=1}^{\infty}$ and the set of NE
decreases to zero. Theoretically, it is possible that multiple limit
points exist for such sequence, hence this convergence behavior is
weaker than the ``strong convergence", in which the sequence
$\{\mathbf{p}^t\}_{t=1}^{\infty}$ admits a single limit point in the
set of NE. In practice though, convergence of the sequence
$\{\mathbf{p}^t\}_{t=1}^{\infty}$ is always observed \footnote{The
S-IWF algorithm proposed in \cite{yu04} for vector MAC channel also
converges to the {\it set} of optimum points similarly as ours, and
in practice it has been observed that such algorithm always
converges to a single point.}. However, for the sake of rigor, in
the next subsection we propose a third algorithm which converges
{\it strongly} to the set of NE.

\subsection{Proposed Algorithm based on Gradient Descent: Strong Convergence}

{\bf Algorithm 3: Projected Gradient Descent Algorithm}:\\
{ In each iteration $t$, the CUs do the
following.\\
1) Calculate the gradient of the potential function:
\begin{align}
\triangledown_{\mathbf{p}_i}
P(\mathbf{p}^t)=\Big[\frac{|h_i(1)|^2}{n(1)+\sum_{j=1}^{N}|h_j(1)|^2
p_j^t(1)},\cdots,\frac{|h_i(K)|^2}{n(K)+\sum_{j=1}^{N}|h_j(K)|^2
p_j^t(K)}\Big]^{\intercal}.
\end{align}

2) Adjust their power profiles simultaneously according to:
\begin{align}
\mathbf{p}^{t+1}_{i}&=\left[\mathbf{p}^{t}_{i}+\alpha_t\triangledown_{\mathbf{p}_i}
P(\mathbf{p}^t)\right]_{\mathcal{P}_i}\triangleq
\bfPsi_i(\mathbf{p}^t)
\end{align}
where the sequence $\{\alpha_t\}_{t=1}^{\infty}$ satisfy
$\alpha_t\in(0,1)$ and \eqref{eqAlphaProperty}; the operator
$[.]_{\mathcal{P}_i}$ represents the projection on to the space
$\mathcal{P}_i$.}

Clearly, this algorithm is based on the classical projected gradient
descent algorithm for solving nonlinear optimization problem, but
with diminishing stepsize $\alpha_t$. In order to prove the
convergence of this algorithm, we first introduce the notion of {\it
Quasi-Fej\'{e}r convergence} \cite{Ermoliev69}, \cite{iusem94},
\cite{Burachik95}.

\newtheorem{D1}{Definition}
\begin{D1}\label{defFejer}
{\it A sequence $\{y^t\}$ is Quasi-Fej\'{e}r convergent to a set
$U\subseteq \mathbf{R}^n$ if for every $u\in U$ there is a sequence
$\{\epsilon_t\}$ such that
$
\epsilon_t\ge 0, ~~\sum^{\infty}_{t=0}\epsilon_t<\infty
$
and
$
||y^{t+1}-u||^2\le ||y^t-u||^2 +\epsilon_t,~\forall~t.
$
}
\end{D1}

The Quasi-Fej\'{e}r sequence has the following property
\cite{iusem94}, \cite{Burachik95}.
\newtheorem{T5}{Theorem}
\begin{T1}\label{theoremFejer}
{\it If $\{y^t\}$ is Quasi-Fej\'{e}r convergent to a nonempty set
$U\subseteq \mathbf{R}^n$, then $\{y^t\}$ is bounded. If furthermore
a limit point $y^*$ of $\{y^t\}$ belongs to $U$, then
$\lim_{k\to\infty}y^k=y^*$, i.e., the sequence converges to a single
point in $U$.}
\end{T1}

Using the notion of Quasi-Fej\'{e}r convergence, we have the
following strong convergence result for Algorithm 3. Please see
Appendix \ref{appTheoremDescent} for proof.
\newtheorem{P4}{Proposition}
\begin{P1}\label{PropDescent}
{\it The projected gradient descent algorithm is Quasi-Fej\'{e}r
convergent to the set of NE of game $G$, with error term
$\epsilon_t\triangleq
{2}\alpha_t\left(\bfPsi(\mathbf{p}^t)-\mathbf{p}^t\right)^{\intercal}
\triangledown_{\mathbf{p}}P(\mathbf{p}^t)$. Moreover, the sequence
$\{\mathbf{p}^t\}_{t=1}^{\infty}$ generated by this algorithm
converges to a point in the set of NE.}
\end{P1}

\subsection{Discussion}\label{subsecDiscussion}

We first note that all the three algorithms proposed in the previous
subsections can be carried out in a distributed fashion. That is, in
order to carry out the computations in each iteration (mainly to
compute $\bfPhi_i(.)$ or $\bfPsi_i(.)$) of the algorithms, the CUs
do not need to know the behavior of other CUs in the network.
Instead, an individual CUs $i$ only needs to know the aggregated
{\it interference plus noise} (IPN) contributed by all other CUs on
each channel: $IPN_i(k)\triangleq n(k)+\sum_{j\ne i}|h_j(k)|^2
p_j(k),~\forall~k\in\mathcal{K}$. As suggested by \cite{Meshkati06},
this information can be fed back to the CUs by the AP. In fact, the
AP  only needs to {\it broadcast} the quantity
$\left\{n(k)+\sum_{i\in\mathcal{N}}|h_j(k)|^2
p_j(k)\right\}_{k\in\mathcal{K}}$ to the CUs, and individual CU $i$
can subtract its contribution and calculate
$\{IPN_i(k)\}_{k\in\mathcal{K}}$. We can also show that, similarly
as in the previous two subsections, that a more general case of the
algorithm where each CU $i$ adopts different sequences of update
coefficients (say $\{\alpha^i_t\}_{t=1}^{\infty}$) also converges,
as long as each sequence $\{\alpha^i_t\}_{t=1}^{\infty}$ satisfies
the conditions in \eqref{eqAlphaProperty}.

As stated previously, the theoretical categorization of the
algorithms by their convergence behaviors is necessary, because it
is generally not possible for the game $G$ to have a single
equilibrium point. Although for the algorithms in both categories,
the potential function (or equivalently the sum capacity) converges
to the single optimum point, the convergence behavior of the
underlying CUs' strategies are more involved. Simply claiming the
algorithm to be ``convergent" might be too ambiguous and sometimes
misleading \footnote{Indeed, in many situations convergence to a set
leads to oscillation of the sequence. For example, the sequence
$\{\frac{1}{t}+\sin(0.5\pi t)\}_{t=1}^{\infty}$ converges to the set
\{-1,~0,~1\}. }. We observe that many iterative water-filling based
algorithms for calculation of the capacity for vector MAC and
broadcast channels, for example the algorithms in \cite{Li07},
\cite{yu04} and \cite{jindal05}, can only be theoretically proven to
be weakly convergent (in which the optimum capacity is attained in
the limit, but the underlying sequence converges to the optimum
set), although in practice they generally converges to a single
optimum point.

For the descent algorithm, note that if the update step size is a
constant, then the algorithm is also weakly convergent (see Prop.
3.4 of \cite{bertsekas97})\footnote{Consequently, most algorithms
proposed for potential games based on projected gradient methods
(e.g., those in \cite{scutari06}) can also be categorized as weak
convergence when the potential function is concave but not strictly
concave.}. The descent algorithm with diminishing step size is also
used in \cite{zhang08} for network utility maximization with
feedback uncertainty, and the problem considered is very different
from ours. We remark that, {\it strong convergence} does not imply
{\it fast convergence}. Indeed, although we are able to show that
the projected gradient decent algorithm converges strongly (which is
theoretically appealing), in practice it tends to converge much
slower than A-IWF and S-IWF. As such, in the second part of this
paper, we will only choose A-IWF and S-IWF as building blocks for
the joint AP selection and power allocation algorithm.


\section{Simulation Results}\label{secSimulation}
In this section, we demonstrate the performance of the proposed
algorithm. We have the following general settings for the
simulation. We place multiple CUs and the AP randomly in a
$10m\times10m$ area; we let $d_{i,w}$ denote the distance between CU
$i$ and AP $w$, then the channel gains between CU $i$ and AP $w$.
Unless otherwise noted,  $\{|h_{i,w}(k)|^2\}_{k\in\mathcal{K}_w}$
are independently drawn from an exponential distribution with mean
$\frac{1}{d^2_{i,w}}$ (i.e., $|h_{i,w}(k)|$ is assumed to have
Rayleigh distribution).

\begin{figure}[htb]
{\includegraphics[width=
0.8\linewidth]{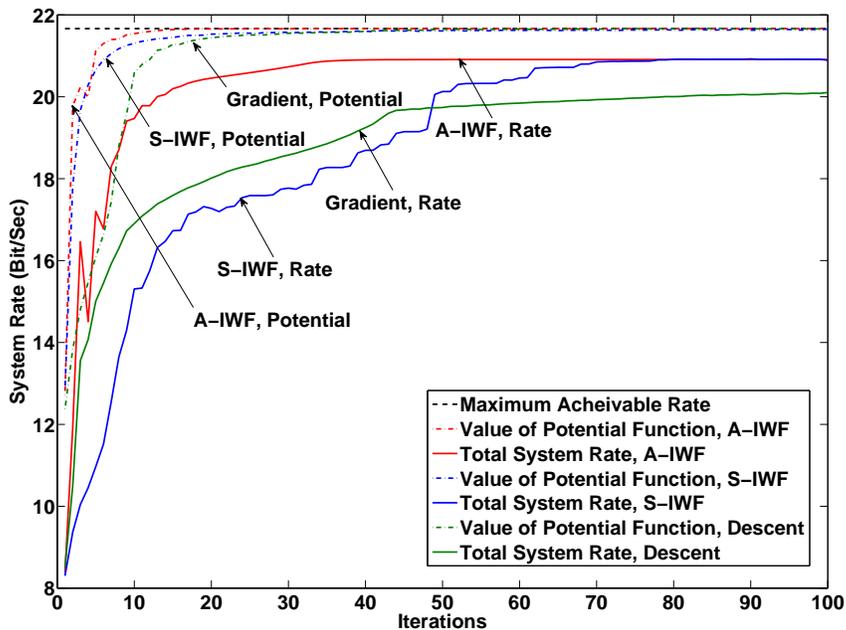}}
\vspace*{-.3cm}\caption{A particular realization of the algorithms
with K=32 and N=10. }\label{figConvergence} \vspace*{-.3cm}
\end{figure}

Fig. \ref{figConvergence} shows a typical realization of the three
algorithms analyzed in this paper, in a network with $10$ CUs and
$32$ channels. It is seen that the values of the potential function
generated by these algorithms converge to the maximum system
capacity quickly, but the sum rate of the CUs (hence individual
power profiles) converges slowly for the projected gradient descent
algorithm.

\begin{figure*}[htb] \vspace*{-.1cm}
    \begin{minipage}[t]{0.5\linewidth}
    \centering
    {\includegraphics[width=
1\linewidth]{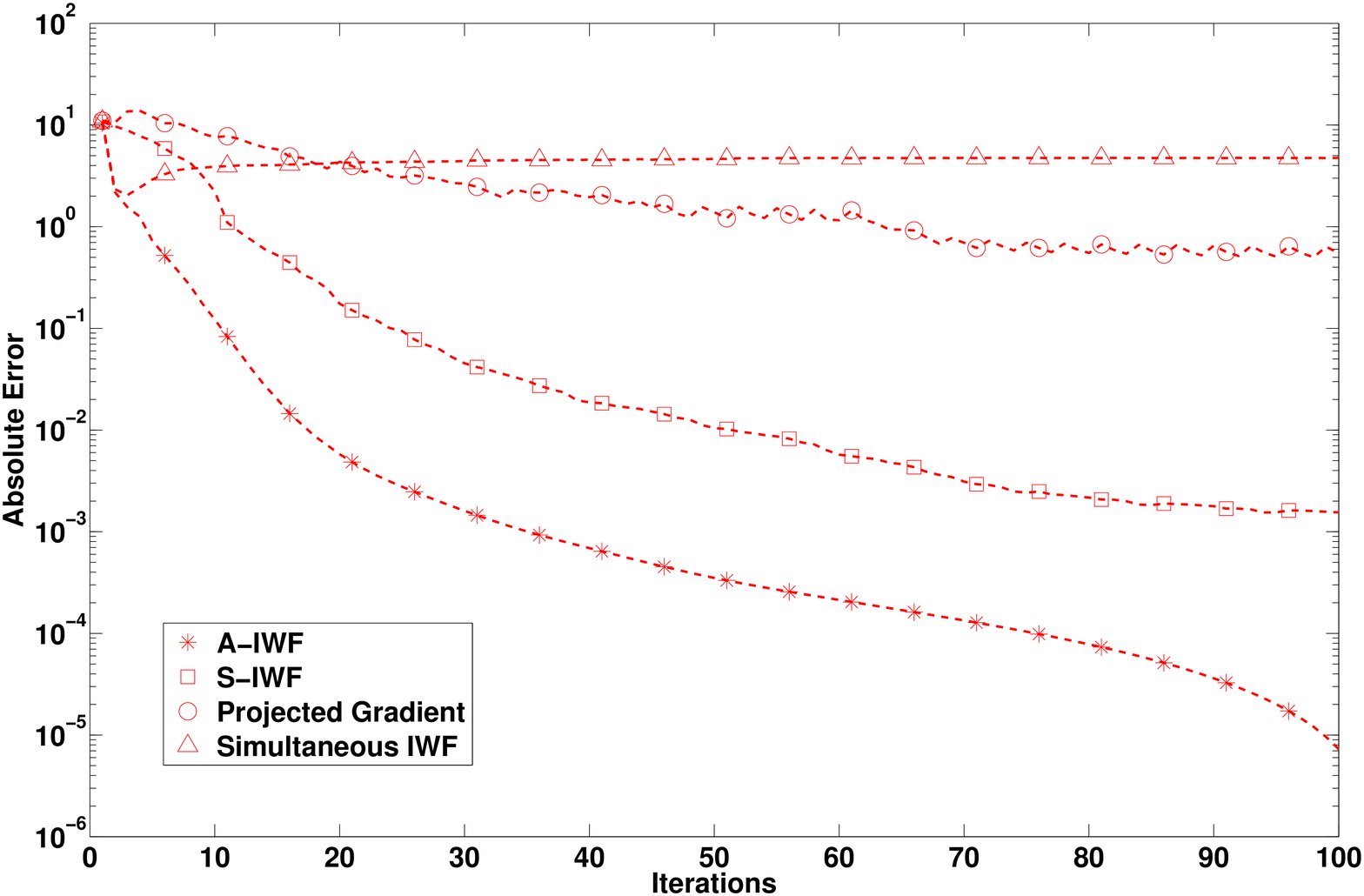}
\vspace*{-0.3cm}\caption{Averaged absolute difference between the
maximum system capacity and the value of potential function
generated by different algorithms. K=32,
N=10.}\label{figConvergenceSpeedABS} \vspace*{-0.3cm}}
\end{minipage}
    \begin{minipage}[t]{0.5\linewidth}
    \centering
    {\includegraphics[width=
1\linewidth]{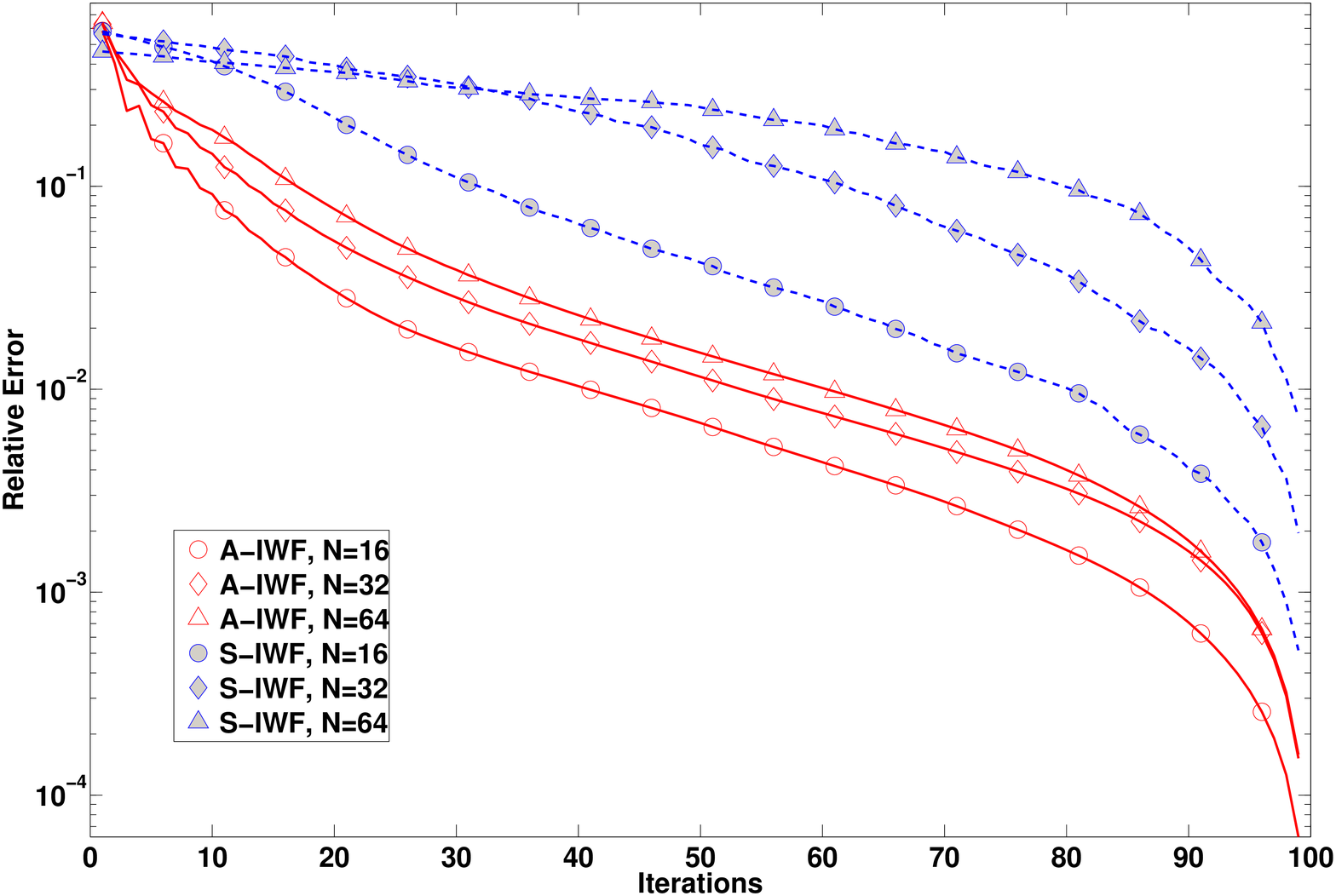}
\vspace*{-0.3cm}\caption{Convergence behavior of the sum rate of CUs
generated by A-IWF and S-IWF with
K=128.}\label{figConvergenceSpeedRelative} \vspace*{-0.3cm}}
\end{minipage}
    \end{figure*}

Fig. \ref{figConvergenceSpeedABS} partly quantifies the convergence
speed of different algorithms. In this figure, we compare the
absolute difference between the maximum system sum rate and the
values of the potential function generated by different algorithms
(i.e., $\{|P(\mathbf{p}^t)-P^*|\}$), in a network with $20$ user and
$64$ channels. We observe that both the A-IWF and S-IWF algorithms
converge relatively fast while the projected gradient descent
algorithm, as seen in Fig.\ref{figConvergence}, converges slowly. We
have also studied the performance of simultaneous IWF algorithm
\cite{scutari08b}, which clearly diverges in our single AP network.
Such phenomenon has been partially explained in Section
\ref{subInapplicableIWF}. 
In Fig. \ref{figConvergenceSpeedRelative}, we characterize the
convergence behavior of the sum of the CUs' rate
$R(t)\triangleq\sum_{i\in\mathcal{N}}R_i(\mathbf{p}^t)$, by plotting
the relative difference between $R(t)$ and $R(100)$:
$\frac{|R(100)-R(t)|}{R(100)}$. Such metric can be viewed as related
to the convergence speed of the algorithm. We see that for network
with 128 channels and with increasing number of CUs, S-IWF converges
increasingly slowly. Such behavior of the S-IWF is intuitively
considering the sequential nature of the algorithm. We note that
each point in both of these two figures is an average of 100
independent runs of the respective algorithms.
%

\begin{figure*}[htb] \vspace*{-.1cm}
    \begin{minipage}[t]{0.5\linewidth}
    \centering
    {\includegraphics[width=
1\linewidth]{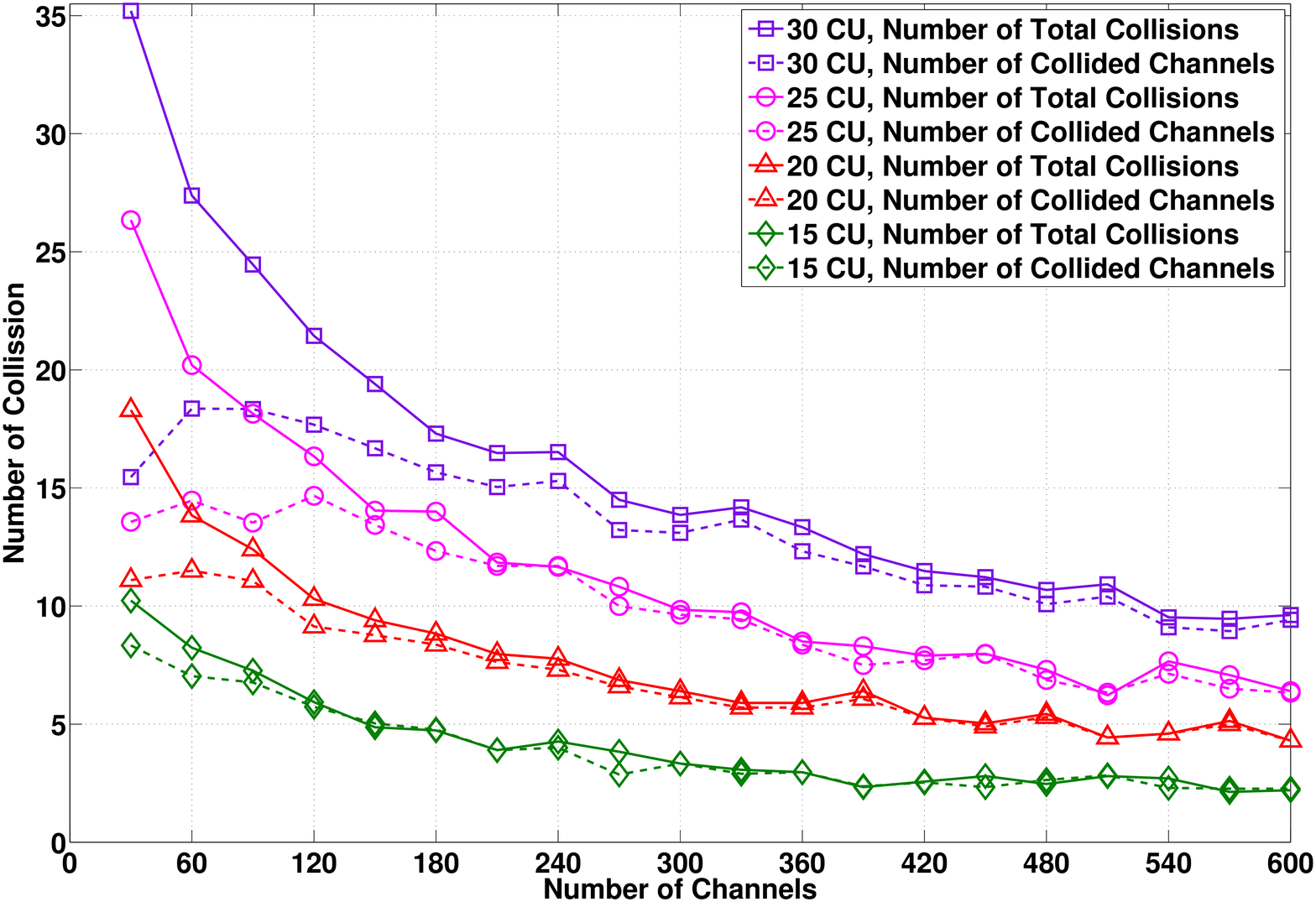}
\vspace*{-0.3cm}\caption{Averaged number of collisions and averaged
number of collided channels.}\label{figCollision} \vspace*{-0.3cm}}
\end{minipage}
    \begin{minipage}[t]{0.5\linewidth}
    \centering
    {\includegraphics[width=
1\linewidth]{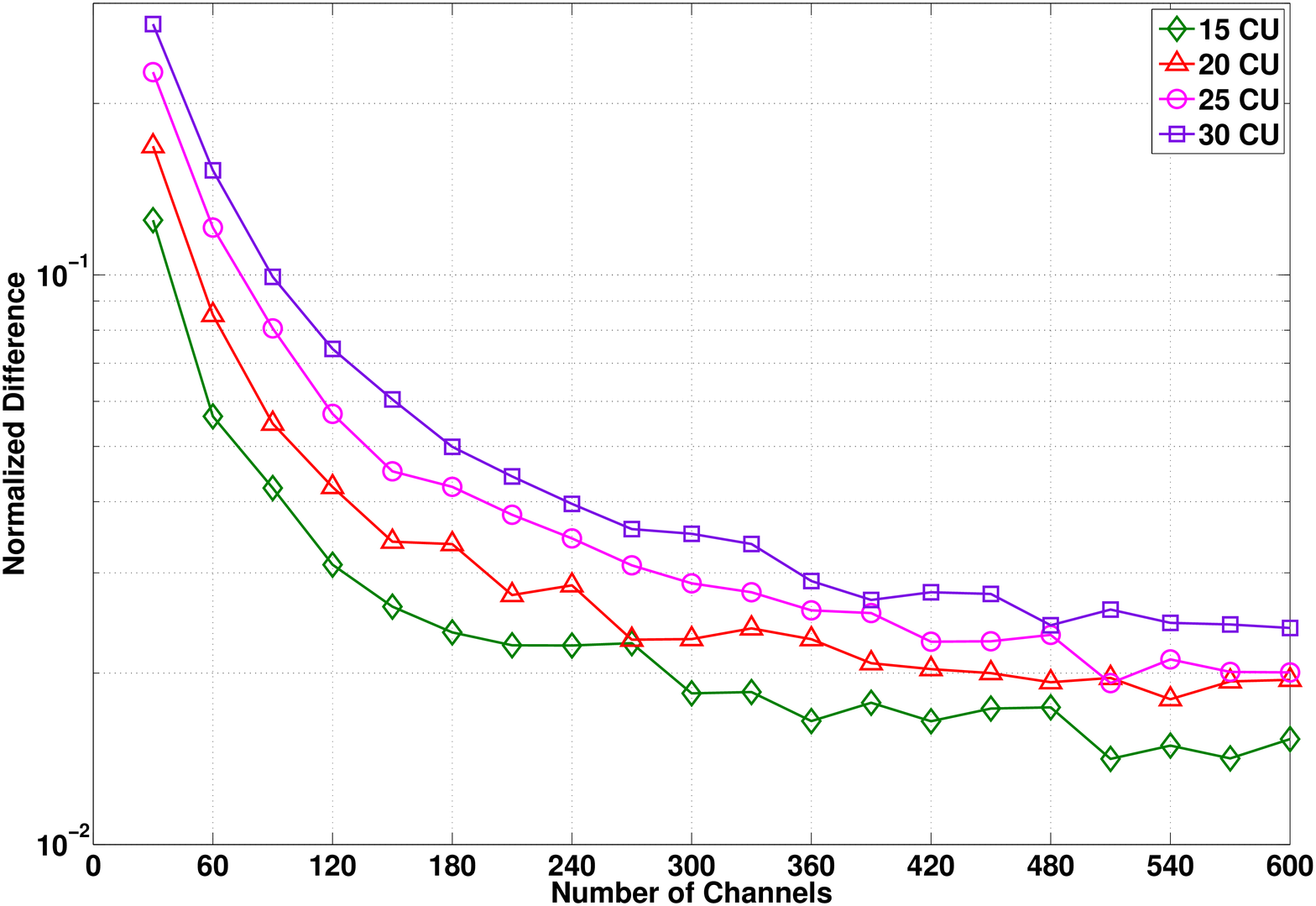}
\vspace*{-0.3cm}\caption{Averaged relative difference between the
sum rate of the CUs and the maximum sum rate of the
system.}\label{figDifferenceWithOptimumCapacity} \vspace*{-0.3cm}}
\end{minipage}
    \end{figure*}

In Section \ref{subNEAsymptotic}, we have predicted that for a fixed
number of CUs, when the number of channels becomes large, the CUs
tend to share the spectrum in a FDMA fashion, and the sum rate of
the users approaches the maximum achievable system sum rate. Fig.
\ref{figCollision} and Fig. \ref{figDifferenceWithOptimumCapacity}
justify these claims. We say that a channel is collided if more than
one CUs are using this channel. We say that a (event of) collision
occurs if two CUs are using the same channel\footnote{If $n\ge2$ CUs
are using the same channel, then there are a total number of
$\frac{(n-1)(n)}{2}$ collisions occurred.}. In Fig.
\ref{figCollision}, we plot the relationship between the number of
channels in the system and the number of collided channels as well
as the total number of collisions. Clearly, as the number of
channels becomes large, both of the above quantities decreases. We
also observe that when the number of channels becomes large, the
number of collided channels tends to be the same as the total number
of collisions, a phenomenon which implies that there tend to be no
more than two CUs using a collided channel. In Fig.
\ref{figDifferenceWithOptimumCapacity}, we show the relative
difference between the sum rate of the CUs after $200$ iteration of
the A-IWF algorithm and the maximum sum rate (i.e.,
$\frac{|R(200)-P^*|}{P^*}$), when the number of channels becomes
large. The decreasing of such relative difference is an indication
of increased efficiency of the spectrum sharing among the CUs. We
note that each point in both of these two figures is again an
average of 100 independent runs of the respective algorithms.

%

\begin{figure*}[htb] \vspace*{-.1cm}
    \begin{minipage}[t]{0.5\linewidth}
    \centering
    {\includegraphics[width=
1\linewidth]{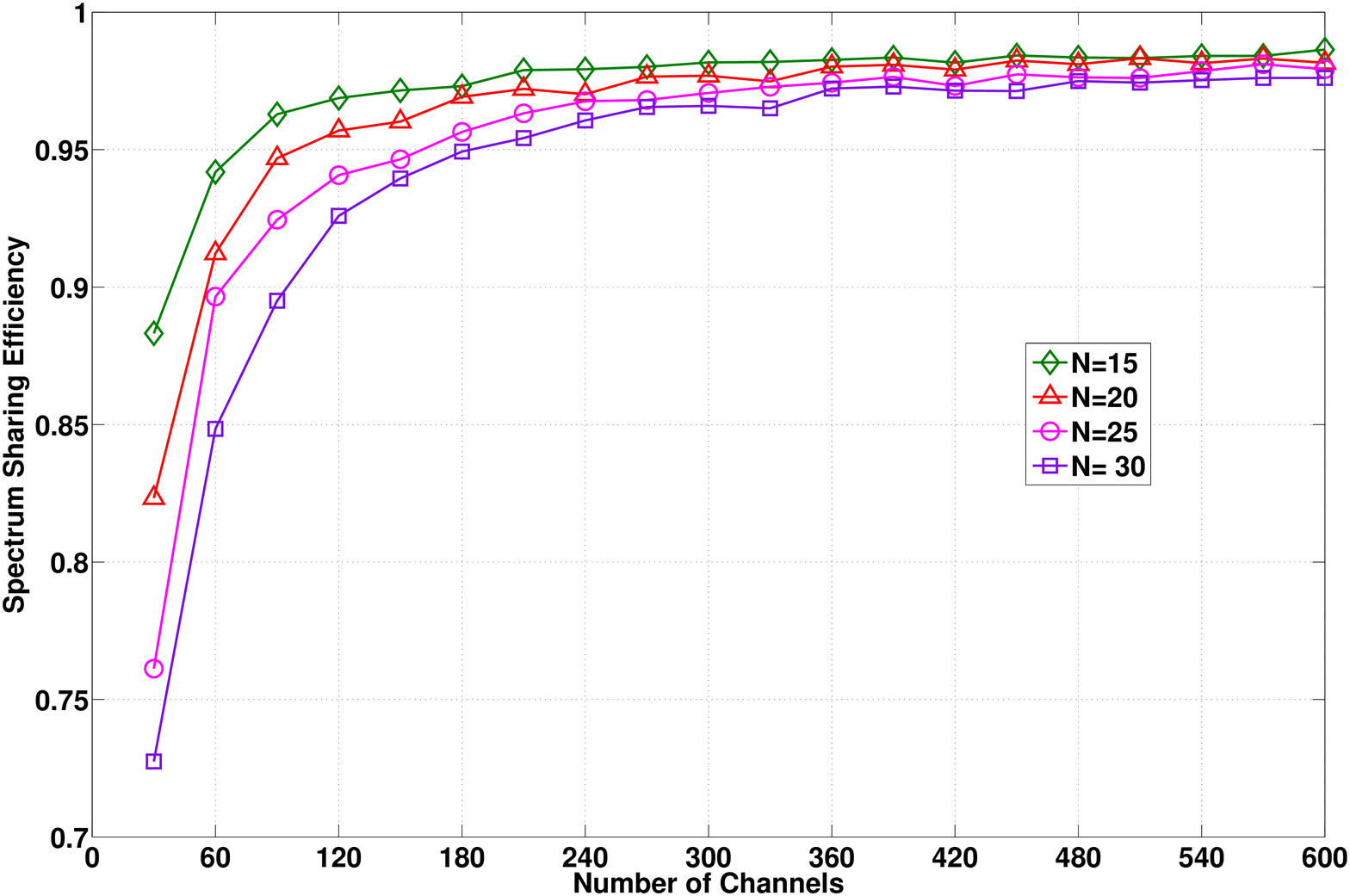}
\vspace*{-0.3cm}\caption{Comparison of the averaged spectrum sharing
efficiency to the number of channels,
N=[15,20,25,30].}\label{figEfficiency} \vspace*{-0.3cm}}
\end{minipage}
    \begin{minipage}[t]{0.5\linewidth}
    \centering
    {\includegraphics[width=
1\linewidth]{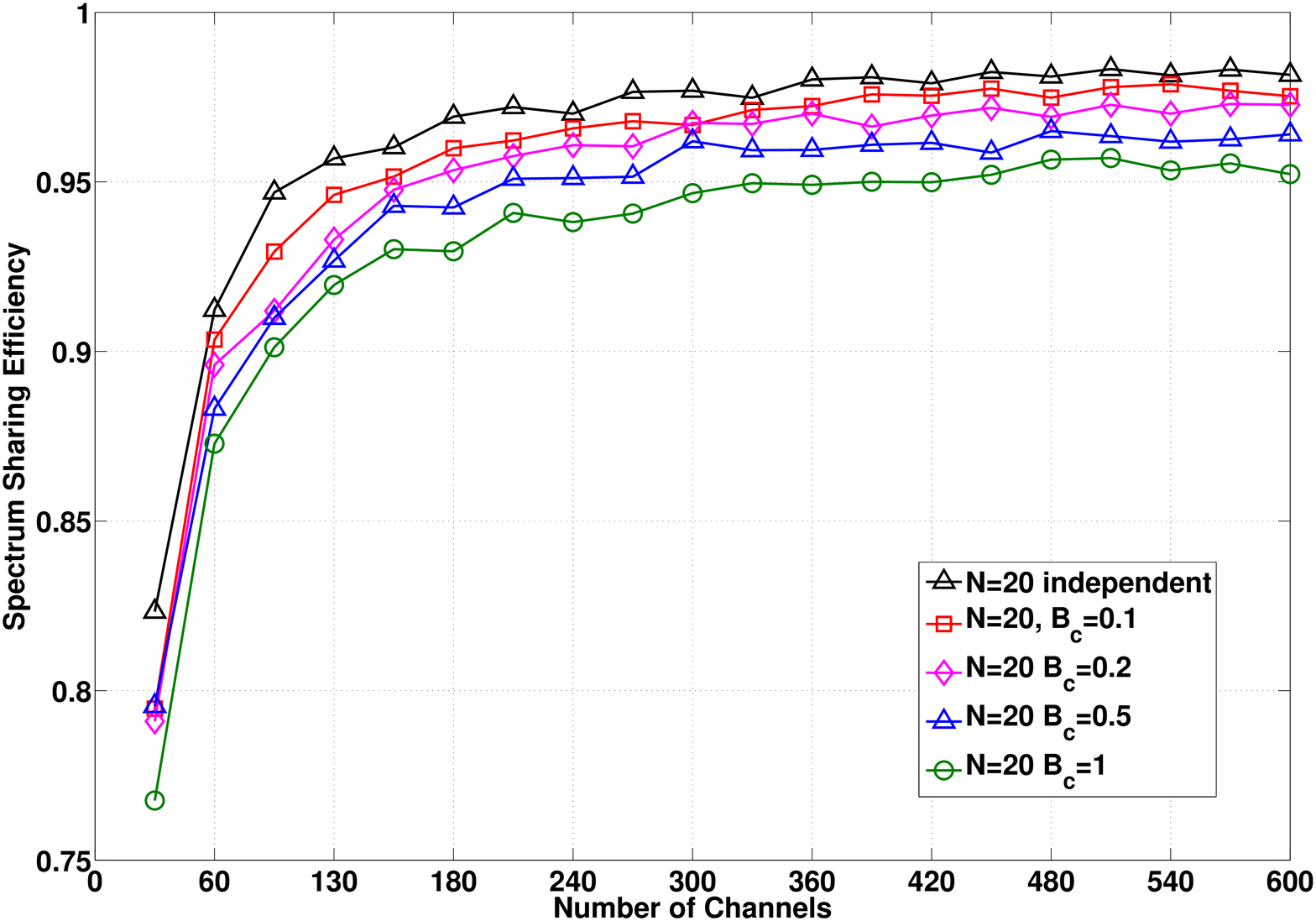}
\vspace*{-0.3cm}\caption{Comparison of the averaged spectrum sharing
efficiencies when the coherent bandwidth is different.
N=20.}\label{figEfficiencyCorrelation} \vspace*{-0.3cm}}
\end{minipage}
    \end{figure*}

To quantify the overall efficiency of the spectrum sharing scheme,
we plot the normalized system sum rate $\frac{R(200)}{P^*}$ in Fig.
\ref{figEfficiency} for the network with different number of CUs and
different number of channels. Clearly the sharing scheme becomes
more efficient when the number of channels becomes large. Notice,
that in all the previous simulation experiments, we assume that the
channel coefficients $\{h_i(k)\}_{k\in\mathcal{K}}$ of a particular
CU $i$ to be independent. This is true when the width of each
channel is comparable to the coherent bandwidth, denoted as $B_c$
\cite{goldsmith05}. However, when we divide a fixed spectrum band
with arbitrarily large number of channels, the coherent bandwidth
eventually becomes larger than the channel width. Indeed, as
mentioned in \cite{Li07}, in practice the parallel frequency
selective channels are usually correlated. As a result, in Fig
.\ref{figEfficiencyCorrelation} we study the spectrum sharing
efficiency for a network with $N=20$ CUs and with networks of
different channel coherent bandwidth $B_c=\{1, ~0.5, ~0.2, ~0.1\}$
(recall that our total available bandwidth is normalized to $1$).
For reference we also plot the case where the channels are assumed
to be independent. We observe that large coherent bandwidth reduces
the sharing efficiency. This phenomenon can be explained by noticing
that when the channel becomes correlated, the event of collision is
more likely to happen, as shown in Table \ref{tableCollision}. We
again note that each point in both of these two figures and each
entry in the table is an average of 100 independent runs of the
respective algorithms.

%

{
\begin{table*}
\caption{Averaged Number of Collisions for Different Channel
Coherent Bandwidth $B_c$ with $N=20$.}
\begin{center}
\small{
\begin{tabular}{|c |c | c | c | c | c |}
\hline
 &Independent & $B_c=0.1$&$B_c=0.2$ & $B_c=0.5$& $B_c=1$ \\
 \hline
 \hline
 Total Collision, K=300 &6.40 & 7.91& 11.12 &  11.60& 14.69\\
 \hline
 Total Collided Channels, K=300 &6.13 & 7.70& 10.06& 10.63 &13.90\\
 \hline
 \hline
 Total Collision, K=600 &4.31 & 5.21 & 6.00 &  7.65&12.67\\
 \hline
 Total Collided Channels, K=600 &4.30 & 5.07& 5.81& 7.61 &12.19\\
\hline
\end{tabular}
} \label{tableCollision}
\end{center}
\vspace*{-0.3cm}
\end{table*}}

\section{Conclusion}\label{secConclusion}
In this first part of the paper, we formulate the uplink spectrum
sharing problem in a single AP CRN into a non-cooperative game
framework. We identify that this game belongs to the family of games
called the ``potential games", and we characterize the properties of
the proposed game. We then propose three algorithms with different
convergence properties that allows the CUs in the network to access
the spectrum in a distributed fashion. From simulation we see that
the proposed algorithms are able to reach the equilibria of the
spectrum sharing game, which represent a set of efficient spectrum
sharing strategies.

In the next part of the paper, we will study jointly the spectrum
sharing and spectrum decision problem in a CRN with multiple APs. We
will see how the algorithms developed in this part of the paper can
be used for constructing efficient and distributed joint spectrum
decision and spectrum sharing strategies.

\appendices
\section{Proof of Lemma
\ref{lemmaAscend}}\label{appTheoremAIWF}
\begin{proof}
We first prove Lemma \ref{lemmaAscend}. We need to show that the
following is true:{\small
\begin{align}
\sum_{i=1}^{N}\sum_{k=1}^{K}\triangledown_{p_i(k)}P(\mathbf{p})s^k_i(\mathbf{p})\ge
\sum_{i=1}^{N}\sum_{k=1}^{K}\left(s^k_i(\mathbf{p})\right)^2 M
\end{align}}
where $s^k_i(\mathbf{p})\triangleq
\Phi^k_i(\mathbf{p}_{-i})-p_i(k)$. It is sufficient to show that for
all $i\in\mathcal{N}$, there must exist a constant $0<M_i<\infty$
such that: $
\sum_{k=1}^{K}\triangledown_{p_i(k)}P(\mathbf{p})s^k_i(\mathbf{p})\ge
\sum_{k=1}^{K}\left(s^k_i(\mathbf{p})\right)^2 M_i.$ In the
following, we will set out to prove that for all $i\in\mathcal{N}$,
there must exist a $M_i$ with  $0<M_i<\infty$, such that:{\small
\begin{align}
\sum_{k=1}^{K}\left(\triangledown_{p_i(k)}P(\mathbf{p})-M_i
s^k_i(\mathbf{p})\right)s^k_i(\mathbf{p})\ge 0.\label{eqPMinorS}
\end{align}}
We notice that{\small
\begin{align}
\triangledown_{p_i(k)}P(\mathbf{p})&=\frac{|h_i(k)|^2}{n(k)+\sum_{j=1}^{N}|h_j(k)|^2p_j(k)},\label{eqPotentialGradient}\\
s^k_i(\mathbf{p})&=\left[\frac{1}{\sigma_i}-\frac{n(k)+\sum_{j\ne
i}|h_j(k)|^2p_j(k)}{|h_i(k)|^2}\right]^{p_{mask}(k)}_{0}\hspace{-0.3cm}-p_i(k)\label{eqSik}.
\end{align}}
We also observe the following equality:{\small
\begin{align}
\sum_{k=1}^{K}s^k_i(\mathbf{p})=\sum_{k=1}^{K}\Phi^k_i(\mathbf{p}_{-i})-\sum_{k=1}^{K}p_i(k)
=\bar{p}_i-\bar{p}_i=0. \label{eqSumSZero}
\end{align}}
This can be readily concluded from our previous observation that
from each user's point of view, it is beneficial to allocate all its
power for communication. Using \eqref{eqSumSZero}, we see that in
order to prove \eqref{eqPMinorS}, it is sufficient to prove that for
all $k_1\in\mathcal{K}$ and $k_2\in\mathcal{K}$ such that{\small
\begin{align}
s^{k_1}_i(\mathbf{p})>0,~~\textrm{and}~~s^{k_2}_i(\mathbf{p})<0\label{eqS1S2}
\end{align}}
there exists $0<M^{(k_1,k_2)}_i<\infty$ such that:{\small
\begin{align}
&\triangledown_{p_i(k_1)}P(\mathbf{p})-M^{(k1,k2)}_i
s^{k_1}_i(\mathbf{p})\ge\triangledown_{p_i(k_2)}P(\mathbf{p})-M^{(k1,k2)}_i
s^{k_2}_i(\mathbf{p})\label{eqK1K2Inequality}\\
\Longleftrightarrow&\triangledown_{p_i(k_1)}P(\mathbf{p})-\triangledown_{p_i(k_2)}P(\mathbf{p})\ge
M^{(k1,k2)}_i \left(s^{k_1}_i(\mathbf{p})-
s^{k_2}_i(\mathbf{p})\right)>0.
\end{align}}
If the above is true, we can take $M_i=\min_{k_1,k_2}M^{(k1,k2)}_i$,
then for all $k_1,~k_2$ that satisfies \eqref{eqS1S2}, we
have{\small
\begin{align}
&\triangledown_{p_i(k_1)}P(\mathbf{p})-\triangledown_{p_i(k_2)}P(\mathbf{p})\ge
M_i \left(s^{k_1}_i(\mathbf{p})-
s^{k_2}_i(\mathbf{p})\right)\nonumber \\
\Longrightarrow&\triangledown_{p_i(k_1)}P(\mathbf{p})-M_i
s^{k_1}_i(\mathbf{p})\ge\triangledown_{p_i(k_2)}P(\mathbf{p})-M_i
s^{k_2}_i(\mathbf{p}).
\end{align}}
Consequently, \eqref{eqPMinorS} can be established.

Let us look at the term $s^{k_1}_i(\mathbf{p})$ first. Let us
simplify the notation by denoting
$s^{k_1}_i(\mathbf{p})=[A_i^{k_1}]^{p_{mask}(k_1)}_{0}-p_i(k_1)$,
where $A_i^{k_1}\triangleq\frac{1}{\sigma_i}-\frac{n(k_1)+\sum_{j\ne
i}|h_j(k_1)|^2p_j(k_1)}{|h_i(k_1)|^2}$. Because
$s^{k_1}_i(\mathbf{p})>0$, we must have that $A_i^{k_1}>p_i(k_1)\ge
0$, consequently, we have:{\small
\begin{align}
0<[A_i^{k_1}]^{p_{mask}(k_1)}_{0}-p_i(k_1)\le
A_i^{k_1}-p_i(k_1).\label{eqK1Inequality}
\end{align}}
We then look at the term $s^{k_2}_i(\mathbf{p})$. We can, similarly
as above, also simplify it as
$s^{k_2}_i(\mathbf{p})=[A_i^{k_2}]^{p_{mask}(k_2)}_{0}-p_i(k_2)$.
Because $s^{k_2}_i(\mathbf{p})<0$, we must have that
$A_i^{k_2}<p_i(k_2)\le p_{mask}(k_2)$, consequently, we have:{\small
\begin{align}
0>[A_i^{k_2}]^{p_{mask}(k_2)}_{0}-p_i(k_2)\ge
A_i^{k_2}-p_i(k_2).\label{eqK2Inequality}
\end{align}}
As a result of \eqref{eqK1Inequality} and \eqref{eqK2Inequality}, in
order to prove \eqref{eqK1K2Inequality}, it is sufficient to prove
that there exists $0<M^{(k1,k2)}_i<\infty$ such that:{\small
\begin{align}
&\triangledown_{p_i(k_1)}P(\mathbf{p})-M^{(k1,k2)}_i\times
(A_i^{k_1}-p_i(k_1))\nonumber\\
&\ge\triangledown_{p_i(k_2)}P(\mathbf{p})-M^{(k1,k2)}_i\times
(A_i^{k_2}-p_i(k_2))\label{eqA1A2Inequality}.
\end{align}}
We see that \eqref{eqA1A2Inequality} is equivalent to{\small
\begin{align}
&\frac{|h_i(k_1)|^2}{n(k_1)+\sum_{j=1}^{N}|h_j(k_1)|^2p_j(k_1)}-
M^{(k1,k2)}_i\left(\frac{1}{\sigma_i}-\frac{n(k_1)+\sum_{j=1}^{N}|h_j(k_1)|^2p_j(k_1)}{|h_i(k_1)|^2}\right)\nonumber\\
&\ge\frac{|h_i(k_2)|^2}{n(k_2)+\sum_{j=1}^{N}|h_j(k_2)|^2p_j(k_2)}-
M^{(k1,k2)}_i\left(\frac{1}{\sigma_i}-\frac{n(k_2)+\sum_{j=1}^{N}|h_j(k_2)|^2p_j(k_2)}{|h_i(k_2)|^2}\right).
\end{align}}
Define
$B_i^{k_1}=\frac{|h_i(k_1)|^2}{n(k_1)+\sum_{j=1}^{N}|h_j(k_1)|^2p_j(k_1)}$,
and $B_i^{k_2}$ similarly, we have that the above inequality can be
simplified to:{\small
\begin{align}
&B_i^{k_1}-M^{(k1,k2)}_i\left(\frac{1}{\sigma_i}-\frac{1}{B_i^{k_1}}\right)\ge
B_i^{k_2}-M^{(k1,k2)}_i\left(\frac{1}{\sigma_i}-\frac{1}{B_i^{k_2}}\right)\nonumber\\
&\Longleftrightarrow
B_i^{k_1}-B_i^{k_2}\ge\left(\frac{1}{B_i^{k_2}}-\frac{1}{B_i^{k_1}}\right)M^{(k1,k2)}_i\label{eqB1B2}.
\end{align}}
Now notice that:{\small
\begin{align}
-\left(A_i^{k_1}-p_i(k_1)\right)+\frac{1}{\sigma_i}=\frac{1}{B_i^{k_1}}~~\textrm{and}~~-\left(A_i^{k_2}-p_i(k_2)\right)+\frac{1}{\sigma_i}=\frac{1}{B_i^{k_2}}
\end{align}}
and we have from \eqref{eqK1Inequality} and \eqref{eqK2Inequality}
that{\small
\begin{align}
-\left(A_i^{k_2}-p_i^{k_2}\right)&\ge
-s^{k_2}_i(\mathbf{p})>0\nonumber\\
0>-s^{k_1}_i(\mathbf{p})&\ge-\left(A_i^{k_1}-p_i^{k_1}\right).
\end{align}}
We have that $ \frac{1}{B_i^{k_1}}<\frac{1}{B_i^{k_2}} \textrm{~ and
~} B_i^{k_1}>B_i^{k_2}$. Consequently, \eqref{eqB1B2} is equivalent
to{\small
\begin{align}
M^{(k_1,k_2)}_i\le
\frac{B_i^{k_1}-B_i^{k_2}}{\frac{1}{B_i^{k_2}}-\frac{1}{B_i^{k_1}}}=B_i^{k_2}\times
B_i^{k_1}.
\end{align}}
Now it is clear that we can always find such a
$0<M^{(k_1,k_2)}_i<\infty$, that satisfies the above inequality,
because the fact that
$B_i^k=\frac{|h_i(k)|^2}{n(k)+\sum_{j=1}^{N}|h_j(k)|^2p_j(k)}$ is
always bounded above and strictly greater than 0
($|h_i(k)|^2>0~\forall~i\in\mathcal{N}~k\in\mathcal{K}$,
$n(k)>0~\forall~k\in\mathcal{K}$).

Now that we can always find $0<M^{(k_1,k_2)}_i<\infty$ that
satisfies \eqref{eqK1K2Inequality}, from the argument above, we can
see that $0<M_i=\min_{k_1,k_2\in\mathcal{K}}M_i^{(k_1,k_2)}<\infty$
must satisfy \eqref{eqPMinorS}. Thus, take
$M=\min_{i\in\mathcal{N}}M_i$, we have that \eqref{eqM} is true.
Thus, the proposition is proved.
\end{proof}

\section{Proof of Proposition \ref{PropDescent}}\label{appTheoremDescent}
\begin{proof}
The projected gradient algorithm can be written as:
$\mathbf{p}^{t+1}=\left[\mathbf{p}^{t}+\alpha_t\triangledown_{\mathbf{p}}
P(\mathbf{p}^t)\right]_{\mathcal{P}}\triangleq
\bfPsi(\mathbf{p}^t)$, where
$\mathcal{P}\triangleq\prod_{i\in\mathcal{N}}\mathcal{P}_i$. We
first show that at least one limit point of the sequence
$\mathbf{p}^*$ is a NE of the game $G$. From the Projection Theorem
(\cite{bertsekas97} Sec 3.3 Prop. 3.2) we have that:
\begin{align}
\left(\bfPsi(\mathbf{p}^t)-\mathbf{p}^t-\alpha_t\triangledown_{\mathbf{p}}
P(\mathbf{p}^t)\right)^\intercal
\left(\mathbf{p}^t-\bfPsi(\mathbf{p}^t)\right)\ge 0.
\end{align}
Consequently, we have:
\begin{align}
0\le||\bfPsi(\mathbf{p}^t)-\mathbf{p}^t||^2\le\alpha_t
\left(\bfPsi(\mathbf{p}^t)-\mathbf{p}^t\right)^\intercal\triangledown_{\mathbf{p}}P(\mathbf{p}^t)\label{eqAscent}.
\end{align}
Similarly as in \eqref{eqTylor}, we invoke the descent lemma:{\small
\begin{align}
F(\bfPsi(\mathbf{p}^t))&\le
F(\mathbf{p}^t)+(\bfPsi(\mathbf{p}^t)-\mathbf{p}^t)^\intercal\triangledown_{\mathbf{p}}F(\mathbf{p}^t)+
\frac{K}{2}||\bfPsi(\mathbf{p}^t)-\mathbf{p}^t||^2\nonumber\\
&\stackrel{(a)}\le F(\mathbf{p}^t)+
(\frac{K}{2}-\frac{1}{\alpha_t})||\bfPsi(\mathbf{p}^t)-\mathbf{p}^t||^2\nonumber\\
&\stackrel{(b)}\le F(\mathbf{p}^t)+
(\frac{K}{2}{\alpha_t}-1)\alpha_t||\triangledown_{\mathbf{p}}P(\mathbf{p}^t)||^2
\end{align}}
where $(a)$ is from \eqref{eqAscent}; $(b)$ is because of the
non-expansiveness of the projection operator:{\small
\begin{align}
||\bfPsi(\mathbf{p}^t)-\mathbf{p}^t||^2&=\left|\left|\left[\mathbf{p}^t+\alpha_t
\triangledown_{\mathbf{p}}P(\mathbf{p}^t)\right]_{\mathcal{P}}-\left[\mathbf{p}^t\right]_{\mathcal{P}}\right|\right|^2\nonumber\\
&\le \alpha^2_t||\triangledown_{\mathbf{p}}P(\mathbf{p}^t)||^2.
\end{align}}
Thus there must exist a time $T^*$ such that $\forall~t>T^*$,
$F(\bfPsi(\mathbf{p}^t))\le F(\mathbf{p}^t)$. From the fact that the
function $F(\mathbf{p})$ is lower bounded, we must have that the
sequence $F(\mathbf{p}^t)$ converges. An immediate consequence of
this result (cf. equation \eqref{eqSummability}) is that:{\small
\begin{align}
\sum_{t=1}^{\infty}(\bfPsi(\mathbf{p}^t)-\mathbf{p}^t)^\intercal\triangledown_{\mathbf{p}}P(\mathbf{p}^t)<\infty\label{eqEpsilonSummable}.
\end{align}}
Let $\mathbf{p}^*$ be a limit point of the sequence
$\{\mathbf{p}^t\}$, then we must have that
$\bfPsi(\mathbf{p}^*)=\mathbf{p}^*$. This fact combined with the
projection theorem implies that for any $\mathbf{y}\in\mathcal{P}$,
the following is true:{\small
\begin{align}
0&\ge\left(\mathbf{y}-\bfPsi(\mathbf{p}^*)\right)^{\intercal}\left(\mathbf{p}^*+
\alpha_t\triangledown_{\mathbf{p}}P(\mathbf{p}^*)-\bfPsi(\mathbf{p}^*)\right)\nonumber\\
&=\alpha_t\left(\mathbf{y}-\mathbf{p}^*\right)^{\intercal}
\triangledown_{\mathbf{p}}P(\mathbf{p}^*).
\end{align}}
The last inequality shows that
$\mathbf{p}^*\in\arg\max_{\mathbf{p}\in\mathcal{P}}P(\mathbf{p})$,
and consequently, $\mathbf{p}^*$ is a NE of the game $G$.

We then show that the sequence $\{\mathbf{p}^t\}$ is Quasi-Fej\'{e}r
convergent to the set of NE. Using again the Projection Theorem, and
(with a little abuse of notation) take $\mathbf{p}^*$ to be {\it
any} NE solution, we have:{\small
\begin{align}
0&\le
\left(\mathbf{p}^*-\bfPsi(\mathbf{p}^t)\right)^{\intercal}\left(\bfPsi(\mathbf{p}^t)-\mathbf{p}^t-
\alpha_t\triangledown_{\mathbf{p}}P(\mathbf{p}^t)\right)\nonumber\\
&=\left(\mathbf{p}^*-\mathbf{p}^t\right)^{\intercal}\left(\bfPsi(\mathbf{p}^t)-\mathbf{p}^t-
\alpha_t\triangledown_{\mathbf{p}}P(\mathbf{p}^t)\right)\nonumber\\
&+\left(\mathbf{p}^t-\bfPsi(\mathbf{p}^t)\right)^{\intercal}\left(\bfPsi(\mathbf{p}^t)-\mathbf{p}^t-
\alpha_t\triangledown_{\mathbf{p}}P(\mathbf{p}^t)\right).
\end{align}}
This is equivalent to:{\small
\begin{align}
&\left(\mathbf{p}^*-\mathbf{p}^t\right)^{\intercal}\left(\bfPsi(\mathbf{p}^t)-\mathbf{p}^t\right)\nonumber\\
&\ge
\alpha_t\left(\mathbf{p}^*-\mathbf{p}^t\right)^{\intercal}\triangledown_{\mathbf{p}}P(\mathbf{p}^t)+
||\bfPsi(\mathbf{p}^t)-\mathbf{p}^t||^2+\alpha_t\left(\mathbf{p}^t-\bfPsi(\mathbf{p}^t)\right)^{\intercal}
\triangledown_{\mathbf{p}}P(\mathbf{p}^t)\nonumber\\
&\stackrel{(a)}\ge||\bfPsi(\mathbf{p}^t)-\mathbf{p}^t||^2+
\alpha_t\left(\mathbf{p}^t-\bfPsi(\mathbf{p}^t)\right)^{\intercal}\triangledown_{\mathbf{p}}P(\mathbf{p}^t)\label{eqProjection}
\end{align}}
where $(a)$ is because of the fact that $P(\mathbf{p})$ is concave:
$\left(\mathbf{p}^*-\mathbf{p}^t\right)^{\intercal}\triangledown_{\mathbf{p}}P(\mathbf{p}^t)\ge
P(\mathbf{p}^*)-P(\mathbf{p}^t)\ge 0$.  The distance between
$\mathbf{p}^*$ and a arbitrary vector $\mathbf{p}^{t+1}$ can be
expressed as follows:{\small
\begin{align}
&||\mathbf{p}^*-\mathbf{p}^{t+1}||^2\nonumber\\
&=||\mathbf{p}^*-\mathbf{p}^{t}||^2
+
||\mathbf{p}^t-\mathbf{p}^{t+1}||^2-2\left(\mathbf{p}^*-\mathbf{p}^t\right)^{\intercal}\left(\mathbf{p}^{t+1}-\mathbf{p}^t\right)\nonumber\\
&\stackrel{(a)}\le
||\mathbf{p}^*-\mathbf{p}^{t}||^2-||\bfPsi(\mathbf{p}^t)-\mathbf{p}^t||^2+
2\alpha_t\left(\bfPsi(\mathbf{p}^t)-\mathbf{p}^t\right)^{\intercal}\triangledown_{\mathbf{p}}P(\mathbf{p}^t)\nonumber\\
&\le||\mathbf{p}^*-\mathbf{p}^{t}||^2+2\alpha_t
\left(\bfPsi(\mathbf{p}^t)-\mathbf{p}^t\right)^{\intercal}\triangledown_{\mathbf{p}}P(\mathbf{p}^t)
\end{align}}
where $(a)$ is from \eqref{eqProjection} and the definition of that
$\bfPsi(\mathbf{p}^{t})=\mathbf{p}^{t+1}$. Now let us take {\small$
\epsilon_t\triangleq
{2}\alpha_t\left(\bfPsi(\mathbf{p}^t)-\mathbf{p}^t\right)^{\intercal}
\triangledown_{\mathbf{p}}P(\mathbf{p}^t). $} Then we have:$
||\mathbf{p}^*-\mathbf{p}^{t+1}||^2\le||\mathbf{p}^*-\mathbf{p}^{t}||^2+
\epsilon_t$. From \eqref{eqAscent} and \eqref{eqEpsilonSummable} we
conclude $\{\epsilon_t\}^{\infty}_{t=1}$ is non-negative and
summable sequence. Because $\mathbf{p}^*$ is an arbitrary NE point,
from Definition \ref{defFejer} the sequence $\{\mathbf{p}^{t}\}$ is
Quasi-Fej\'{e}r convergent to the set of NE of game $G$. The first
part of this proof show that a limit point of $\{\mathbf{p}^t\}$
belongs to the set of NE, consequently, by applying Theorem
\ref{theoremFejer}, we see that $\{\mathbf{p}^t\}$ converges to a
point in the set of NE.
\end{proof}


\bibliographystyle{IEEEbib}
\bibliography{ref}

\begin{thebibliography}{10}

\bibitem{hong11_infocom}
M.~Hong, A.~Garcia, and J.~Barrera,
\newblock ``Joint distributed {AP} selection and power allocation in cognitive
  radio networks\,''
\newblock in {\em the Proceedings of the IEEE INFOCOM}, 2011,
\newblock accepted.

\bibitem{akyildiz08}
I.~F. Akyildiz, W.~Y. Lee, M.~C. Vuran, and S.~Mohanty,
\newblock ``A survey on spectrum management in cognitive radio networks,''
\newblock {\em IEEE Communications Magazine}, pp. 40--48, April 2008.

\bibitem{lai08}
L.~Lai and H.~E. Gamal,
\newblock ``The water-filling game in fading multiple-access channels,''
\newblock {\em IEEE Transactions on Information Theory}, vol. 54, no. 5, 2008.

\bibitem{Meshkati06}
F.~Meshkati, M.~Chiang, H.~V. Poor, and S.~C. Schwartz,
\newblock ``A game-theoretic approach to energy-efficient power control in
  multicarrier {CDMA} systems,''
\newblock {\em IEEE Journal on Selected Areas in Communications}, vol. 24, pp.
  1115--1129, 2006.

\bibitem{islam08}
M.~H. Islam, Y.-C. Liang, and A.~T. Hoang,
\newblock ``Joint power control and beamforming for cognitive radio networks,''
\newblock {\em IEEE Transactions on Wireless Communications}, vol. 7, no. 7,
  pp. 2415--2419, 2008.

\bibitem{stevenson09}
C.~R. Stevenson, G.~Chouinard, Z.~Lei, W.~Hu, S.~J. Shellhammer, and
  W.~Caldwell,
\newblock ``{IEEE} 802.22: the first cognitive radio wireless regional area
  network standard,''
\newblock {\em Comm. Mag.}, vol. 47, no. 1, pp. 130--138, 2009.

\bibitem{song05a}
G.~Song and Y.~Li,
\newblock ``Cross-layer optimization for {OFDM} wireless networks--part {I}:
  Theoretical framework,''
\newblock {\em IEEE Transactions on Wireless Communications}, vol. 4, no. 2,
  pp. 614--624, 2005.

\bibitem{song05b}
G.~Song and Y.~Li,
\newblock ``Cross-layer optimization for {OFDM} wireless networks--part {II}:
  Algorithm development,''
\newblock {\em IEEE Transactions on Wireless Communications}, vol. 4, no. 2,
  pp. 625--634, 2005.

\bibitem{luo04}
Z-.Q. Luo, T.~N. Davidson, G.~B. Giannakis, and K.~M. Wong,
\newblock ``Transceiver optimization for block-based multiple access through
  {ISI} channels,''
\newblock {\em IEEE Transactions on Signal Processing}, vol. 52, no. 4, pp.
  1037--1052, 2004.

\bibitem{yu02b}
W.~Yu and J.~M. Cioffi,
\newblock ``{FDMA} capacity of gaussian multiple-access channel with isi,''
\newblock {\em IEEE Transactions on Communications}, vol. 50, no. 1, pp.
  102--111, 2002.

\bibitem{kim05b}
K.~Kim, Y.~Han, and S.-L Kim,
\newblock ``Joint subcarrier and power allocation in uplink {OFDMA} systems,''
\newblock {\em IEEE Communication Letters}, vol. 9, pp. 526--528, 2005.

\bibitem{liu10}
T.~Liu, C.~Yang, and L.-L. Yang,
\newblock ``A lower-complexity subcarrier-power allocation scheme for
  frequency-division multiple-access scheme,''
\newblock {\em IEEE Transactions on Wireless Communications}, vol. 11, no. 5,
  pp. 1571--1576, 2010.

\bibitem{Li07}
H.~Li and H.~Liu,
\newblock ``An analysis of uplink {OFDM} optimality,''
\newblock {\em IEEE Transactions on Wireless Communications}, vol. 6, no. 8,
  pp. 2972--2983, 2007.

\bibitem{he08}
G.~He, S.~Gault, M.~Debbah, and E.~Altman,
\newblock ``Distributed power allocation game for uplink ofdm systems,''
\newblock in {\em Proc. WiOPT}, 2008, pp. 515--521.

\bibitem{acharya09}
J.~Acharya and R.~D. Yates,
\newblock ``Dynamic spectrum allocation for uplink users with heterogeneous
  utilities,''
\newblock {\em IEEE Transactions on Wireless Communications}, vol. 8, no. 3,
  pp. 1405--1413, 2009.

\bibitem{yu04}
W.~Yu, W.~Rhee, S.~Boyd, and J.~M. Cioffi,
\newblock ``Iterative water-filling for gaussian vector multiple-access
  channels,''
\newblock {\em IEEE Transactions on Information Theory}, vol. 50, no. 1, pp.
  145--152, 2004.

\bibitem{monderer96}
D.~Monderer and L.~S. Shapley,
\newblock ``Potential games,''
\newblock {\em Games and Economics Behaviour}, vol. 14, pp. 124--143, 1996.

\bibitem{scutari08a}
G.~Scutari, D.~P. Palomar, and S.~Barbarossa,
\newblock ``Optimal linear precoding strategies for wideband noncooperative
  systems based on game theory -- part {I}: Nash equilibria,''
\newblock {\em IEEE Transactions on Signal Processing}, vol. 56, no. 3, 2008.

\bibitem{luo06b}
Z-.~Q. Luo and J-.S. Pang,
\newblock ``Analysis of iterative waterfilling algorithm for multiuser power
  contorl in digital subscriber lines,''
\newblock {\em EURASIP Journal on Applied Signal Processing}, vol. 2006, pp.
  1--10, 2006.

\bibitem{zhao07}
Q.~Zhao and B.~M. Sadler,
\newblock ``A survey of dynamic spectrum access,''
\newblock {\em IEEE Signal Processing Magazine}, , no. 5, pp. 79--89, 2007.

\bibitem{cover05}
T.~M. Cover and J.~A. Thomas,
\newblock {\em Elements of Information Theory, second edition},
\newblock Wiley, 2005.

\bibitem{osborne94}
M.~J. Osborne and A.~Rubinstein,
\newblock {\em A Course in Game Theory},
\newblock MIT Press, 1994.

\bibitem{deb08}
R.~Deb,
\newblock ``A characterization of differentiable potential games,''
\newblock {\em http://www.econ.yale.edu/~rd287/}.

\bibitem{scutari06}
G.~Scutari, S.~Barbarossa, and D.~P. Palomar,
\newblock ``Potential games: A framework for vector power control problems with
  coupled constraints,''
\newblock in {\em the Proceedings of ICASSP 06}, 2006.

\bibitem{cheng93}
R.~Cheng and S.~Verdu,
\newblock ``Gaussian multiaccess channels with isi: Capacity region and
  multiuser water-filling,''
\newblock {\em IEEE Transactions on Information Theory}, vol. 39, no. 3, pp.
  773--785, 1993.

\bibitem{yu02a}
W.~Yu, G.~Ginis, and J.~M. Cioffi,
\newblock ``Distributed multiuser power control for digital subscriber lines,''
\newblock {\em IEEE Journal on Selected Areas in Communications}, vol. 20, no.
  5, pp. 1105--1115, 2002.

\bibitem{scutari08b}
G.~Scutari, D.~P. Palomar, and S.~Barbarossa,
\newblock ``Optimal linear precoding strategies for wideband noncooperative
  systems based on game theory -- part {II}: Algorithms,''
\newblock {\em IEEE Transactions on Signal Processing}, vol. 56, no. 3, 2008.

\bibitem{shum07}
K.~W. Shum, K.~K. Leung, and C.~W. Sung,
\newblock ``Convergence of iterative waterfilling algorithm for gaussian
  interference channels,''
\newblock {\em IEEE Journal on Selected Area in Communications}, vol. 25, pp.
  1091--1100, 2007.

\bibitem{bertsekas97}
D.~P. Bertsekas and J.~N. Tsitsiklis,
\newblock {\em Parallel and Distributed Computation: Numerical Methods, 2nd
  ed},
\newblock Athena Scientific, Belmont, MA, 1997.

\bibitem{Ermoliev69}
Y.~M. Ermoliev,
\newblock ``On the method of generalized stochastic gradient and
  quasi-fej\'{e}r sequences,''
\newblock {\em Cybernetics}, 1969.

\bibitem{iusem94}
A.~N. Iusem, B.~F. Svaiter, and M.~Teboulle,
\newblock ``Entropy-like proximal methods in convex programming,''
\newblock {\em Mathematics of Operations Research}, 1994.

\bibitem{Burachik95}
R.~Burachik, L.~M.~G. Drummond, and A.~N. Iusem,
\newblock ``Full convergence of the steepest descent method with inexact line
  search,''
\newblock {\em Optimization}, pp. 137--146, 1995.

\bibitem{jindal05}
N.~Jindal, W.~Rhee, S.~Vishwanath, S.~A. Jafar, and A.~Goldsmith,
\newblock ``Sum power iterative water-filling for multi-antenna gaussian
  broadcast channels,''
\newblock {\em IEEE Transactions on information theory}, vol. 51, no. 4, 2005.

\bibitem{zhang08}
J.~Zhang, D.~Zhang, and M.~Chiang,
\newblock ``The impact of stochastic noisy feedback on distributed network
  utility maximization,''
\newblock {\em IEEE Transactions on Information Theory}, , no. 2, pp. 645--665,
  2008.

\bibitem{goldsmith05}
A.~Goldsmith,
\newblock {\em Wireless Communications},
\newblock Combridge University Press, New York, 2005.

\end{thebibliography}

\end{document}